\documentclass[aps,prb,twocolumn,showpacs,floatfix,superscriptaddress,floatfix]{revtex4}
\usepackage[english]{babel}
\usepackage{graphicx}
\usepackage{color}
\usepackage[utf8]{inputenc}
\usepackage{pstricks,pst-grad,color}
\usepackage{graphicx,pifont,amssymb}
\usepackage{amssymb}


\begin{document}

\title{Large and Small Fermi-Surface Spin Density Waves in the Kondo
  Lattice Model}

\author{Robert Peters}
\email[]{robert.peters@riken.jp}
\affiliation{Computational Condensed Matter Physics Laboratory, RIKEN, Wako, Saitama 351-0198, Japan}

\author{Norio Kawakami}
\affiliation{Department of Physics, Kyoto University, Kyoto 606-8502,  Japan}

\date{\today}

\begin{abstract}
We demonstrate the existence of metallic spin density waves (SDWs) in the Kondo
lattice model on a square lattice for a wide
range of parameters by means of real space dynamical mean field
theory. In these SDWs, the spin polarization as well as the charge
density depend on the lattice site and are modulated along one
direction of the square lattice.
We show that within this phase of metallic SDWs 
the Fermi surface changes from small to large, when the coupling
strength is increased. Furthermore, the transition between the large
Fermi-surface SDW phase and the
paramagnetic phase is of second order, while the transition between the small
Fermi-surface SDW phase and the paramagnetic phase is of first order.
A local quantum critical point is thus avoided in our calculations by
undergoing a first order phase transition.
\end{abstract}


\maketitle
\section{Introduction}
Strongly correlated materials are of special interest in condensed
matter physics because they often exhibit remarkable phenomena which
cannot be explained by single particle theory.
A particular class of strongly correlated materials are heavy fermion
systems, in which quantum criticality accompanied by non-Fermi liquid
behavior or unconventional superconductivity can often be observed.\cite{Si2001,Si200623,lohneysen2007,coleman2007,coleman2005,gegenwart2008,Knafo2009,Si2010pss,Si2010,Si2013,Paschen2014}
Heavy fermion materials include strongly interacting {\it f}-electrons
which are hybridized with conduction {\it spd}-bands. The main driving
forces in these materials are the Kondo screening, which favors a
paramagnetic state, and 
the Ruderman-Kasuya-Kittel-Yosida (RKKY) interaction, which favors a
magnetically long-range ordered state. Especially, when both effects are of
comparable strength, fascinating many-particle phenomena such as
quantum criticality and non-Fermi-liquid physics can arise.

It has been found that quantum phase transitions in heavy fermion
systems can be categorized into two classes: (i) Quantum
fluctuations at the phase transition arise solely due to the magnetic order and the Fermi
surface behaves 
smoothly across the transition, which corresponds to the Hertz-Millis-Moriya
type of quantum critical point (QCP),\cite{Hertz1976,Millis1993,Moriya1995} and (ii) exactly at the
quantum phase transition between magnetism and paramagnetism, not only
the magnetic order vanishes, but also the Kondo screening vanishes and
the Fermi surface changes abruptly, which is
referred to as local quantum criticality.\cite{Si2001,Coleman2001} While CeNi$_2$Ge$_2$ is an
example for the Hertz-Millis-Moriya type of QCP,\cite{gegenwart2008} there is evidence that
the QCP observed in YbRh$_2$Si$_2$ corresponds to a local quantum
critical point.\cite{Gegenwart2002,Custers2003,Paschen2004,Gegenwart2007,Hartmann2010,Friedemann2010} These experimental and theoretical results on quantum
phase transitions in heavy fermion systems have been summarized in the
global phase diagram\cite{Si200623,Senthil2004,Yamamoto2007,Vojta2008,Yamamoto2008,Yamamoto2010b,Si2010pss,Si2010,Si2013,Si2014} which includes small/large Fermi-surface
antiferromagnetic phases, and small/large Fermi-surface paramagnetic
phases.

A commonly used theoretical model to describe heavy fermion systems is the Kondo
lattice model, in which effective local moments originating from
the interacting 
{\it f}-electrons are coupled to conduction electrons. 
The Kondo lattice model reads\cite{doniach77,lacroix1979,fazekas1991,Tsunetsugu1997,Assaad1999}
\begin{eqnarray}
H&=&t\sum_{<i,j>\sigma}c^\dagger_{i\sigma}c_{j\sigma}+J\sum_i\vec{S}_i\vec{s}_i\label{KLM}\\
\vec{s}_i&=&c^\dagger_{i\sigma_m}\vec{\rho}_{\sigma_m\sigma_n}c_{i\sigma_n},\nonumber
\end{eqnarray}
where $c_{i\sigma}^\dagger$ creates an electron with spin-direction
$\sigma$ on lattice site $i$. The first term in Eq.(\ref{KLM})
describes the hopping of conduction electrons, and the second term
describes the coupling of the conduction electrons to the localized
spins with interaction strength $J$. Throughout this paper we assume
an antiferromagnetic coupling, $J>0$, and take the hopping constant
$t$ as unit of energy. All calculations are performed on a
two-dimensional square lattice.

In this paper we examine magnetism and quantum phase transitions in 
the Kondo lattice model by means of the dynamical mean
field theory (DMFT),\cite{Metzner1989,Pruschke1995,Georges1996} which has
been extensively used for this
purpose.\cite{Burdin2000,Zhu2003,Beach2004,Ohashi2005,Peters2007,Beach2008,Otsuki2009,Otsuki2009b,Hoshino2010,Bodensiek2011,Benlagra2011,Peters2011conf,Peters2012,Bercx2012,Bodensiek2013,Peters2013,Peters2013b,Hoshino2013,Golez2013,Kuramoto2014,Kikuchi2014,Osolin2015,Otsuki2015}
Although these previous studies include calculations for the
antiferromagnetic state and the magnetic quantum phase transition,
they were performed for the N\'eel state with 
commensurate ordering vector. Exactly at half filling for a
particle-hole symmetric lattice, the N\'eel state is insulating. Thus,
the behavior of the Fermi surface across the quantum critical point
cannot be studied. Some of these previous studies also used the N\'eel
state away from half filling within the metallic antiferromagnetic
phase. However, away from 
half filling, one can expect that the N\'eel state becomes
energetically unstable towards incommensurate magnetic states, as has
also been found in many experimental 
systems by neutron scattering, e.g. for
CeRu$_2$Si$_2$,\cite{Regnault1988} CeCu$_2$Si$_2$,\cite{Stockert2004}
CeRhIn$_5$,\cite{Bao2000} etc. A theoretical study using classical
spins already proved the existence of incommensurate SDWs.\cite{Hamada1995} However, due to the approximation of classical
spins, many-particle effects, particularly the Kondo effect, are absent
in Ref. 61.

In this paper, we use the DMFT, and explicitly include the possibility of
incommensurate magnetic states in the Kondo lattice model by using
large-scale real-space DMFT (RDMFT).
We will focus on the analysis of metallic SDW states away from half
filling, which have not been carefully studied yet, and analyze
static and dynamical properties with particular emphasis on
the behavior of the Fermi surface across the quantum phase transition.

This paper is organized as follows: In the next section (II), we will
shortly explain the methods used in this paper. This is followed by the
section (III), in which we will summarize the calculated phase diagram and
explain the different types of SDWs observed in our
calculations. Static (IV), dynamical properties (V) and the phase 
transitions between different phases (VI) are explained in more detail
in the subsequent 
sections, before concluding the paper.

\section{Method}
We use large-scale real-space dynamical mean field theory (RDMFT) to study the Kondo
lattice model, Eq. (\ref{KLM}), defined on a two-dimensional square
lattice. In RDMFT each lattice site of a finite
cluster  is mapped
onto its own impurity model. This mapping is done by calculating the
local Green's function of each lattice site via
\begin{eqnarray}
\mathbf{G}_{loc}(z)=\int dk_xdk_y\left[z\cdot\mathbb{I}-\mathbf{H}_{k_x,k_y}-\mathbf{\Sigma} \right]^{-1},\label{ilocalGreen}
\end{eqnarray}
where $\mathbf{H}_{k_x,k_y}$ is the hopping Hamiltonian of the chosen cluster and
$\mathbf{\Sigma}$ is the self-energy matrix. The
momentum-dependence of $\mathbf{H}_{k_x,k_y}$ arises thereby through
the boundary conditions. For this study, we use $(20\times 20)$-,
$(60\times 10)$- and $(30\times 30)$-cluster with periodic boundary conditions. 
The self-energy is
diagonal and momentum-independent within this approximation;
correlation effects between 
different lattice sites are not included. However, the self-energy
depends on the lattice site of the cluster, which makes it
possible to describe SDW states.
From the local Green's function, a lattice-site-dependent
hybridization is calculated via 
\begin{equation}
\Delta_{ii}(z)=z-\frac{1}{G_{loc,ii}(z)}-\Sigma_{ii}(z)\label{hybridization},
\end{equation}
which determines the input for a single impurity Kondo model. These
equations are reduced to the usual single-site DMFT, if the cluster
consists of a single site. 
To obtain the self-energy of the Kondo impurity model, we
use the numerical renormalization group
(NRG),\cite{wilson1975,bulla2008} which can calculate
reliable self-energies at zero and finite temperature.\cite{peters2006,weichselbaum2007}
We then iterate Eqs. (\ref{ilocalGreen}) and (\ref{hybridization}) until
the self-energy matrix becomes self-consistent.
In order to improve the stability of the iterative calculations, we do
not directly use the self-energy as calculated by the NRG, but we add
$50\%$ of the self-energy of the last iteration,
$\Sigma_{i+1}=0.5\Sigma_i+0.5\Sigma_{i-1}$, which is usually
called ``mixing'' and frequently used in DMFT calculations.
Recently, we have used similar
calculations to study spin- and charge density waves in the Hubbard
model.\cite{Peters2014}

\section{Phase diagram}
\begin{figure}[t]
\begin{center}
\includegraphics[width=\linewidth]{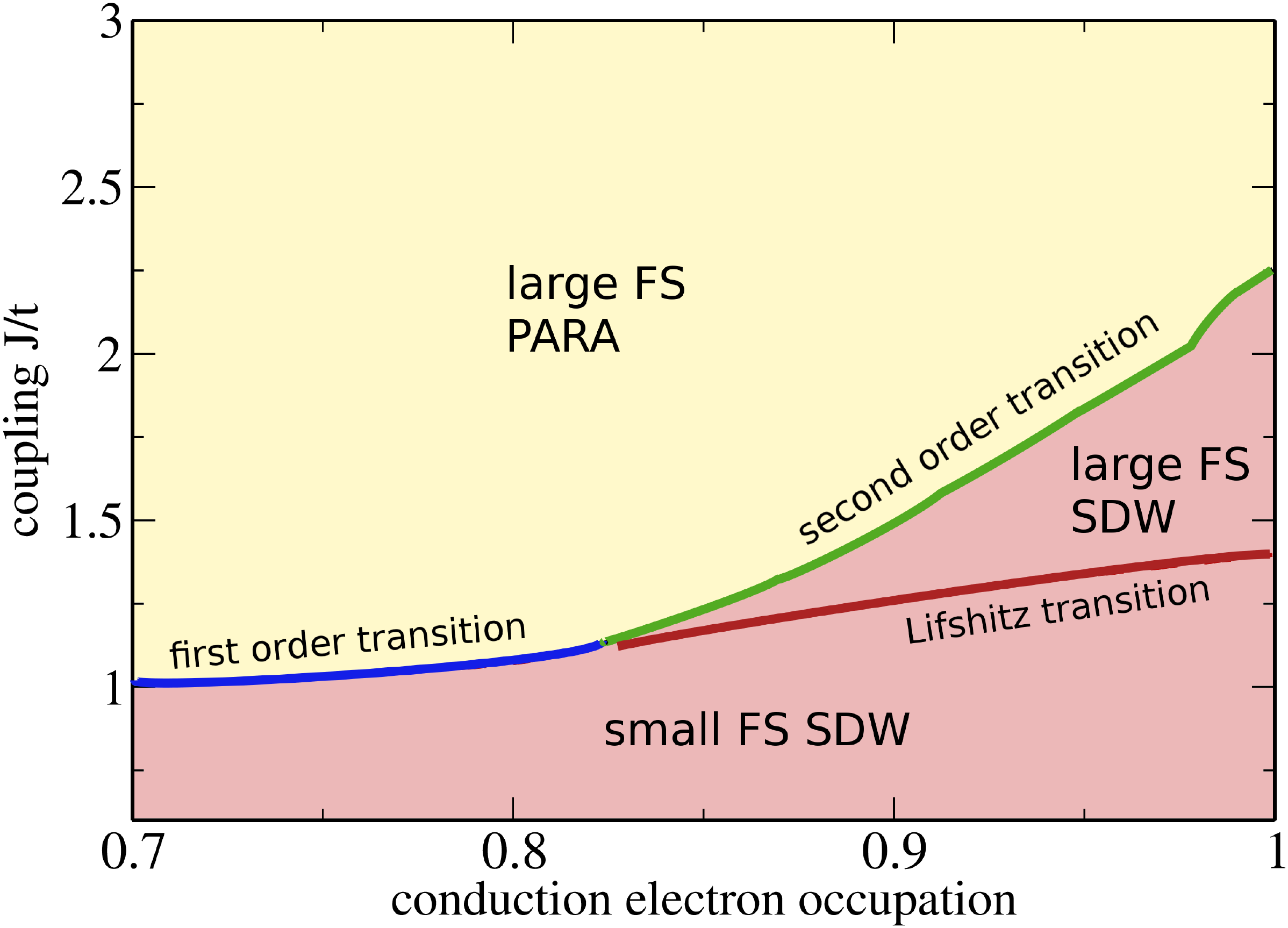}
\end{center}
\caption{Phase diagram of the Kondo lattice model as calculated by
  RDMFT. The SDW phase away from half
  filling is separated into a large Fermi-surface (FS) phase at strong
  coupling and a small Fermi-surface phase at weak coupling. A
  detailed explanation is given in the main text.
\label{phase_diagram}}
\end{figure}
Figure \ref{phase_diagram} summarizes our obtained results. 
Close to half filling, the physics of the Kondo lattice model is
dominated by the interplay of the Kondo effect and the RKKY
interaction, as described by the Doniach phase diagram.\cite{doniach77}
At weak coupling, where the RKKY interaction dominates, we observe a
magnetically ordered phase. At strong coupling, where the
Kondo effect becomes dominating, a paramagnetic state is stabilized.
\begin{figure}
\begin{center}
\includegraphics[width=0.99\linewidth]{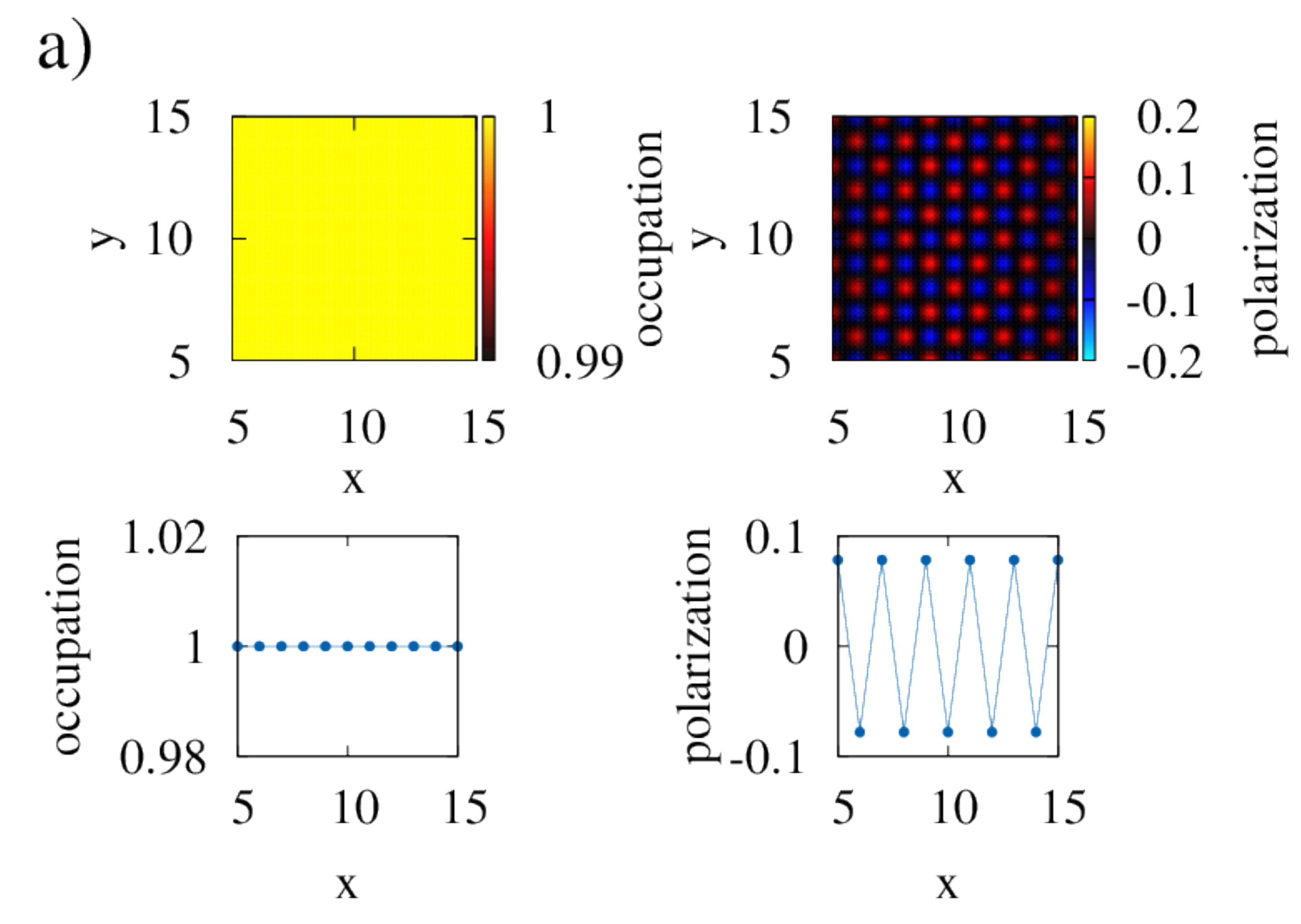}\\
\includegraphics[width=0.99\linewidth]{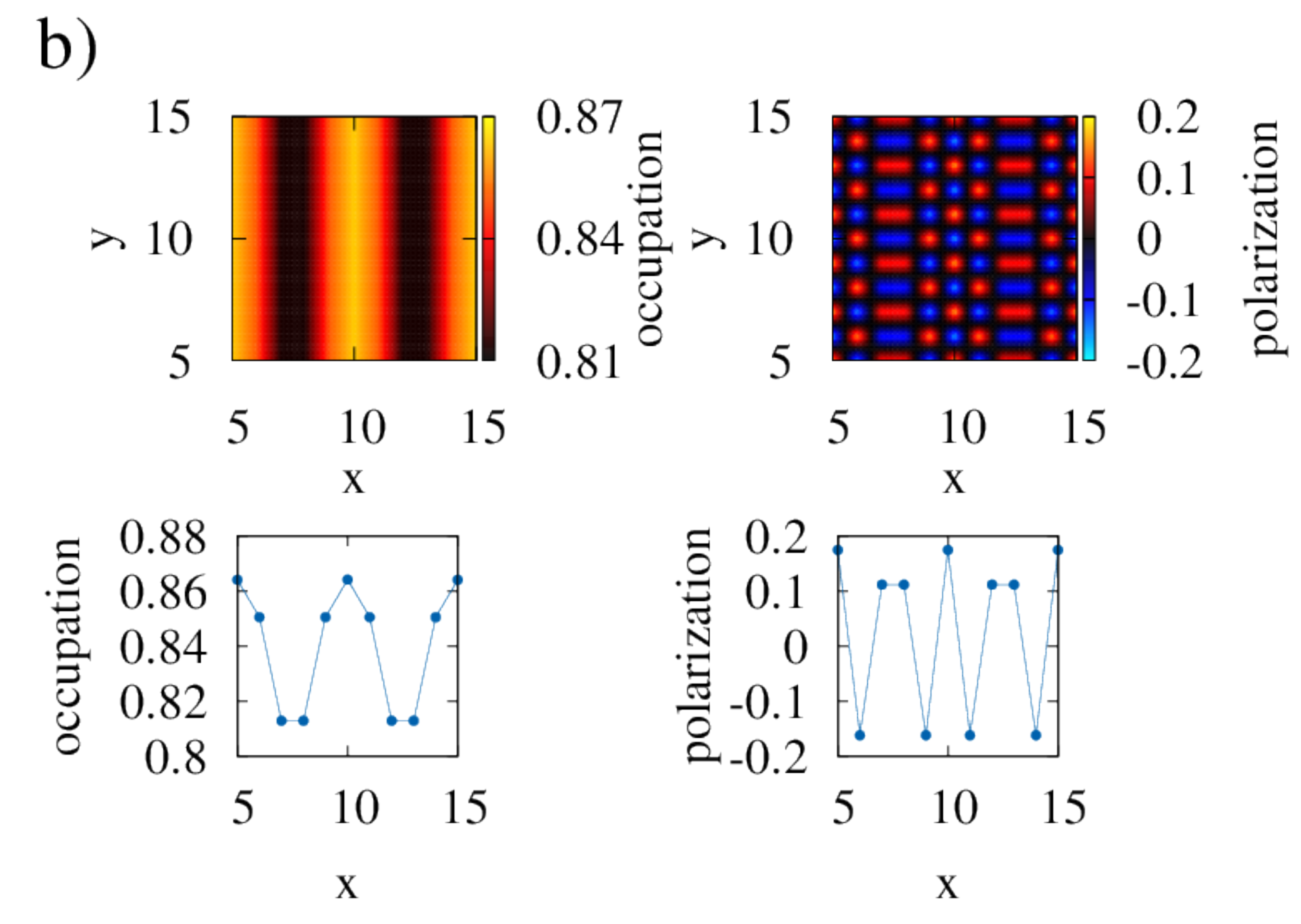}\\
\includegraphics[width=0.99\linewidth]{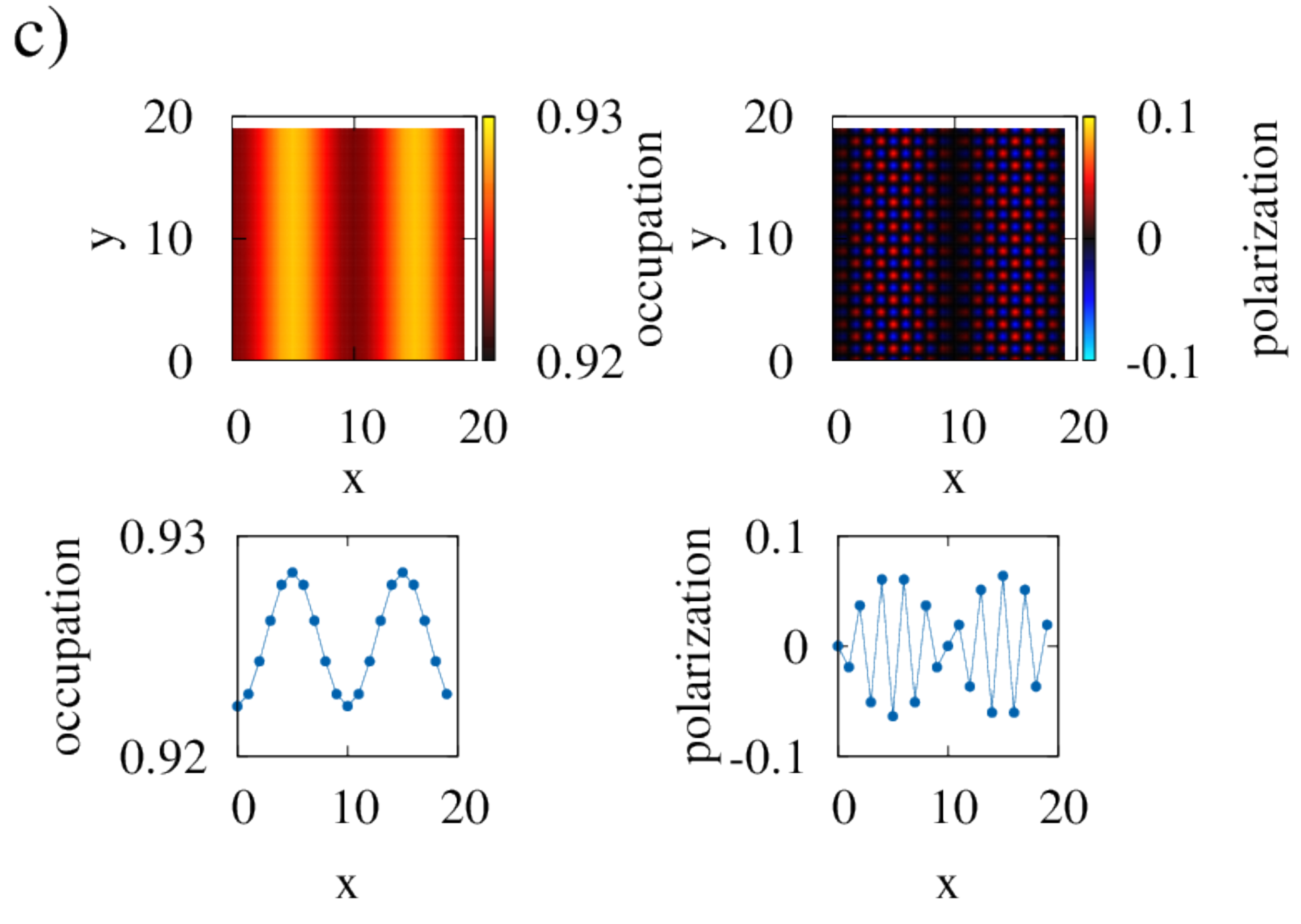}
\end{center}
\caption{Different magnetic states close to half filling: a)
  antiferromagnetic N\'eel state exactly at half filling ($J=2t$). b)
  doped SDW/CDW at weak interaction
  strength ($J=0.8t$). c) doped SDW/CDW close
  to the quantum critical phase transition ($J=1.8t$). For each state a)-c), we
  show a false-color-plot of the charge density (top left), a
  false-color-plot of the electron polarization (top right), an
  intersection of the charge-density plot for a given $y$-site (bottom
  left), and an intersection of the electron-polarization plot for a given $y$-site (bottom right).
\label{AF_states}}
\end{figure}

Exactly at half filling the model exhibits perfect nesting and we
observe the well-known transition from an 
antiferromagnetic N\'eel state at weak coupling to a paramagnetic
Kondo insulator at strong coupling. This continuous transition occurs
approximately at $J_c/t\approx 2.2$ within DMFT for the square lattice. Notice that the Kondo lattice
model on a square 
lattice remains insulating at half filling for all interaction
strengths, $J>0$, although the origin of the insulating gap gradually changes.
Our obtained critical interaction strength for the transition
between the N\'eel state and the Kondo insulator reasonably agrees with
previous DMFT calculations.\cite{Otsuki2015} We
can thus reproduce previous results with our RDMFT calculations.

Away from half filling, the homogeneous N\'eel state becomes unstable
and changes into a
phase of SDWs, in which the polarization of the
conduction electrons as well as the localized spins are lattice-site dependent.
For coupling
strengths larger than the hopping amplitude, this
SDW phase becomes unstable upon doping towards the large Fermi-surface
paramagnetic state.
On the other hand, for weak coupling, $J<t$, SDWs can be observed for
any conduction band filling.

Even within this SDW phase, the modulation of the electron polarization
and electron density depends on the model parameters.
In Fig. \ref{AF_states}, we compile information about 
typical SDWs,
which can be observed in the vicinity of half filling.  
For each type of SDW, a)-c), we show in the upper part of each panel
false-color plots of the electron density (left plot) and the electron
polarization (right plot) for a cluster of lattice sites. The lower
plots of each panel show intersections of the above false-color plots.

Panel a) shows the antiferromagnetic N\'eel
 state exactly at half filling. In the N\'eel state, the polarization
 of the electrons changes its sign between
nearest neighbors. The electron density, on the
other hand, is unity for all lattice sites. Increasing the interaction
strength beyond $J/t=2.2$, the antiferromagnetic state changes into the
 Kondo insulator (not shown in Fig. \ref{AF_states}), where all lattices
 sites are occupied in average with a 
 single electron and the polarization vanishes. 

Away from half filling, the perfect nesting, which exists exactly at
half filling, is lost, and the system can gain energy by modulating the
electron polarization over the lattice sites. We have tried in our calculations
different types of modulations, such as vertical SDWs and diagonal
SDWs. It turned out that the most stable type of SDW on the square
lattice is the vertical SDW, as shown in panel b) and c) of
Fig. \ref{AF_states}. The amplitude of the electron polarization is
modulated along one direction of the square lattice. Along the other
direction, the polarization shows a N\'eel-state-like 'AB'
oscillation. However, not only the polarization is modulated, but also the
electron density depends on the lattice site; there are stripes of
high electron density and stripes of low electron density.
CDWs are well known in strongly correlated materials,
  e.g. High-Tc Cuprates, and have
  been detected there by X-ray scattering.\cite{Abbamonte2005}
 Within the
regions of high electron density, we generally observe a N\'eel-state-like behavior; the polarization changes its sign between nearest
neighbors. The behavior in the low density regions, however, depends
on the interaction parameter and the average occupation of the conduction
electrons.
For large interaction strengths close to the quantum critical
transition, we observe that the polarization vanishes in the low
electron-density regions, see Fig. \ref{AF_states} panel c). The Kondo
effect becomes strong in these 
regions so that the localized moments are screened by the Kondo effect
instead of forming a magnetically ordered state. 
On the other hand, for weak interaction strength, $J<t$, and
especially for an average 
electron filling $n<0.9$, we observe that the low electron-density
regions are still magnetically polarized. We find that nearest
neighbor sites form ferromagnetic bonds along
one axis, see Fig. \ref{AF_states} panel b). 
For weak interaction strengths, the localized moments cannot be
screened by the Kondo effect, which is exponentially weak. 
Here the magnetic state is completely determined by the
  RKKY interaction.  
The localized moments are nearly fully polarized and form also inside
the low electron-density region a magnetic state. Furthermore, the
modulation of the 
  polarization cannot be described anymore by a simple sine-wave, but
  a more complex polarization pattern is realized.
Due to a changed Fermi vector in this region,
the system can gain energy by combining ferromagnetic bonds with
antiferromagnetic ones. Experimentally, similar combinations of
antiferromagnetic and ferromagnetic correlations have been observed in YbRh$_2$Si$_2$.\cite{Ishida2002,Lausberg2013}

\section{static properties of the SDW phase}
\begin{figure}[tb]
\begin{center}
\includegraphics[width=\linewidth]{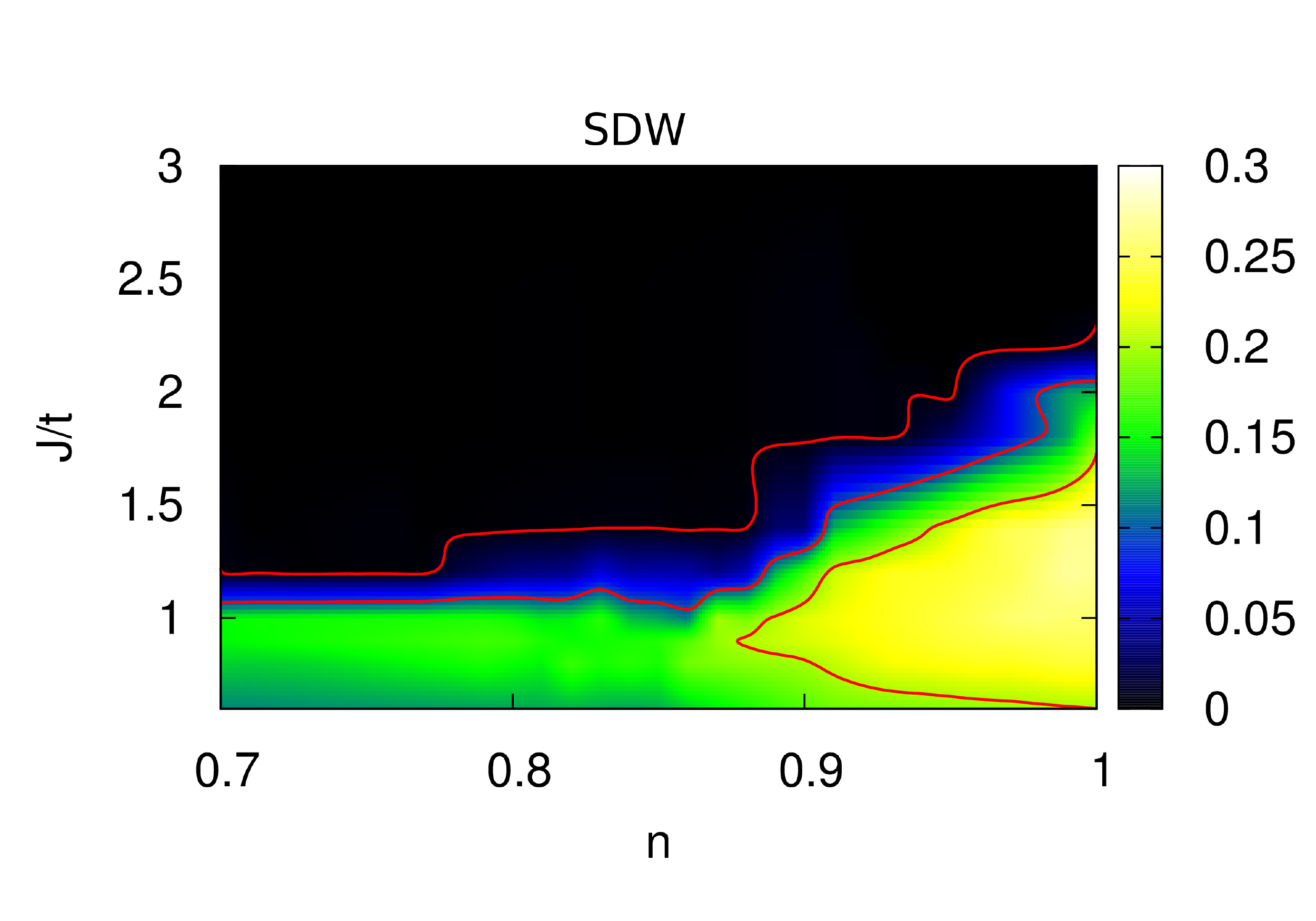} 
\includegraphics[width=\linewidth]{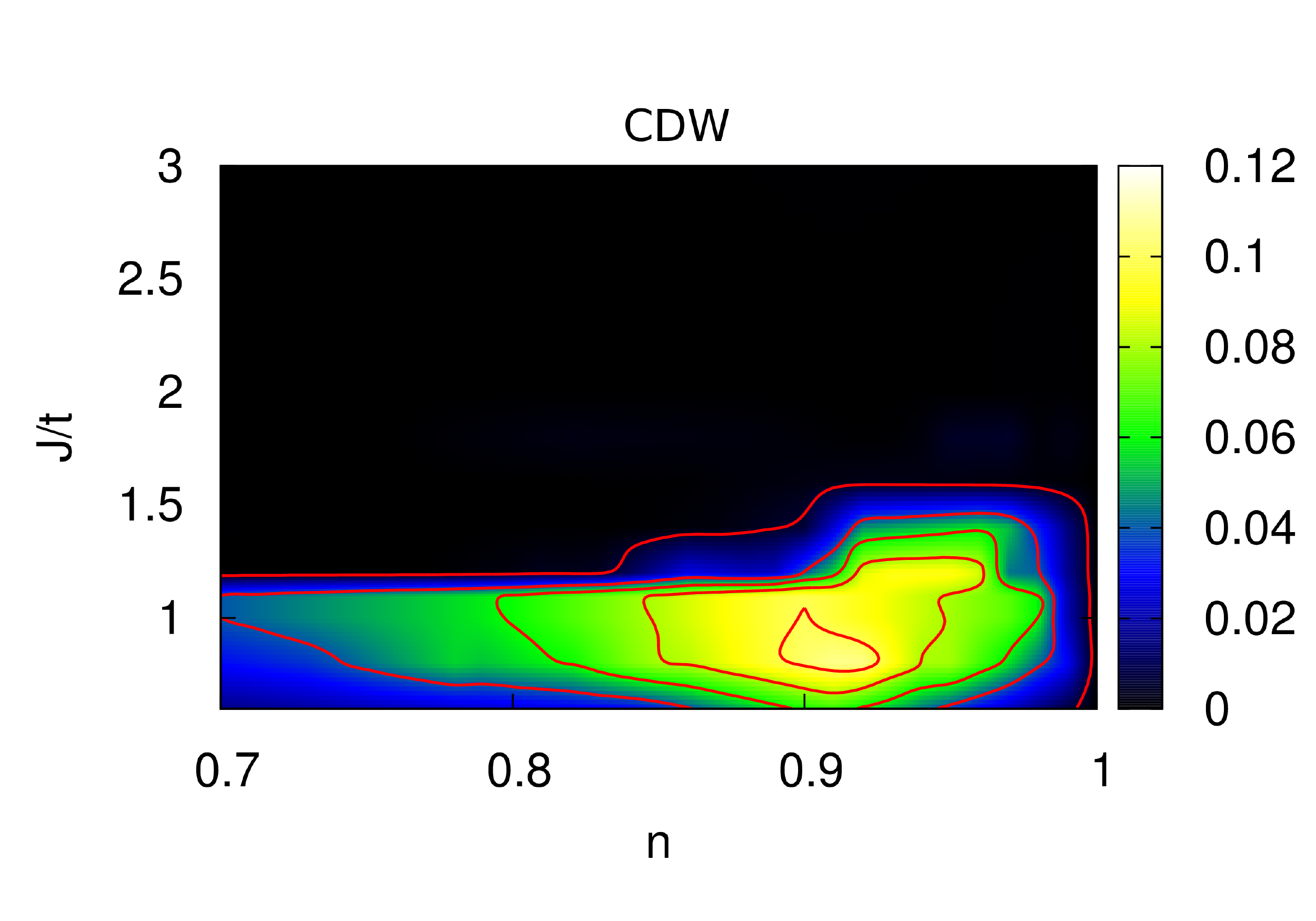} 
\end{center}
\caption{Upper panel: Electron polarization in the high density
  regions (corresponding to the maximal polarized electrons) for different
  coupling strengths and conduction band fillings. The red lines
  correspond to contour lines at $\vert
  n_\uparrow-n_\downarrow\vert=(0$; $0.1$; $0.2)$.
Lower panel: The
  difference of the electron density between high and low density
  regions. The red lines
  correspond to contour lines at $\vert
  n_{\text{high}}-n_{\text{low}}\vert=(0$; $0.03$; $0.06$; $0.09$; $0.12)$.
\label{SDW_CDW}}
\end{figure}
An important point which was neglected in previous studies is that
both the spin polarization and the electron density are
modulated away from half filling.
In Fig. \ref{SDW_CDW}, we show the amplitude of the SDW
(upper panel) and the amplitude of the CDW (lower panel) for different coupling
strengths and conduction band fillings. We observe that the electron
polarization within the SDW is largest for approximately $J\approx t$ exactly at half
filling. The spin polarization becomes weaker when doping the system
away from half filling. 
The amplitude of the CDW, however, is strongest for $J\approx t$
and conduction band fillings $\langle n \rangle\approx 0.9$. Exactly at half
filling, the electron density is homogeneous. Furthermore, the CDW is
coupled to the SDW; within the paramagnetic state, no charge-density
modulation exists. The Kondo effect alone is not sufficient to stabilize a
CDW state for these conduction band fillings.
With increasing coupling strength, the CDW vanishes faster than the
SDW. However, this might be related to the fact that the SDW
state can only be found very close to half filling for strong
interaction. In this parameter region the ground state becomes more
and more homogeneous, and the amplitude of the CDW is very small.

\begin{figure}[tb]
\begin{center}
\includegraphics[width=\linewidth]{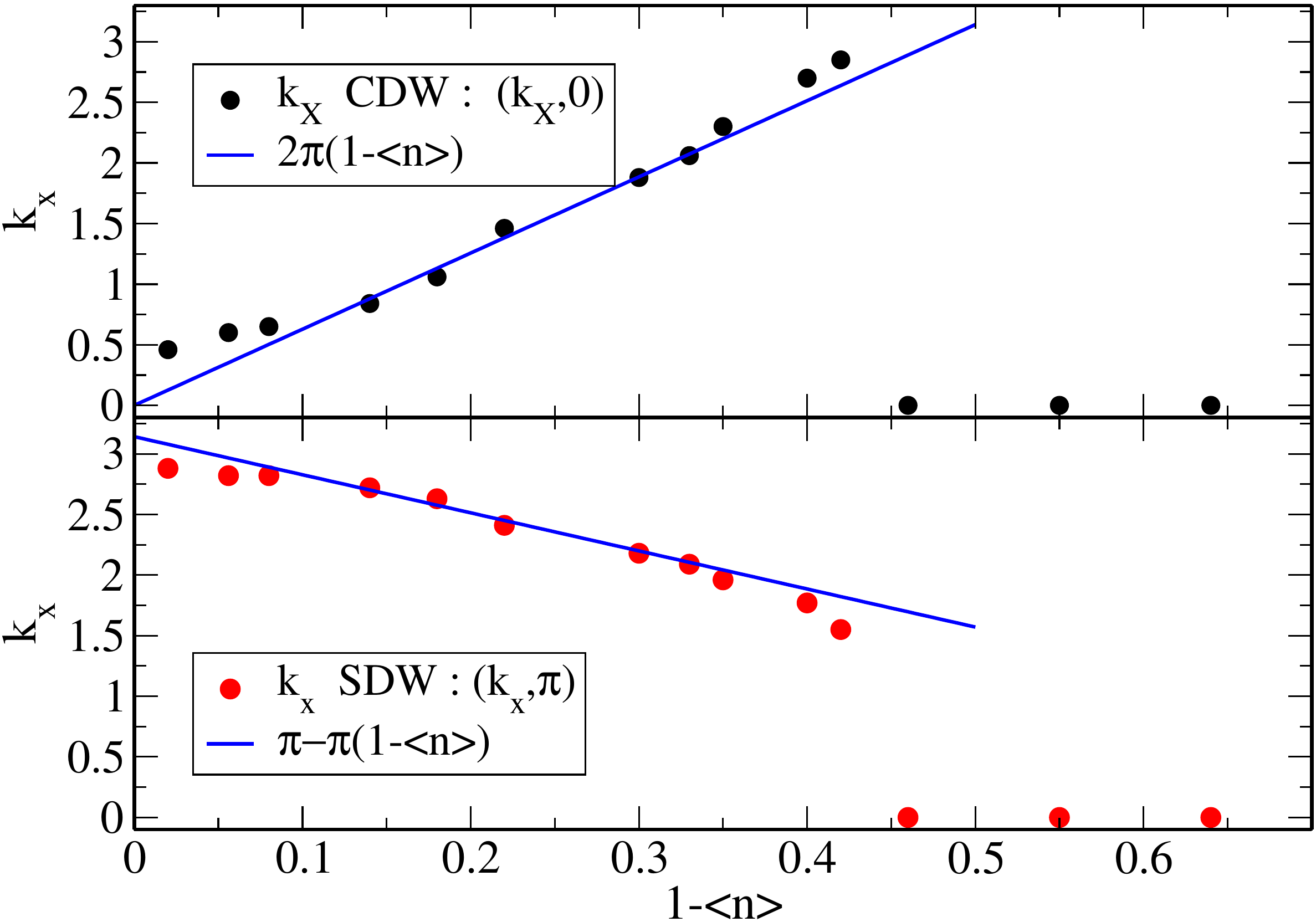}
\end{center}
\caption{Strongest Fourier mode of the charge density
  (upper panel) and spin polarization (lower panel) for different
  conduction band fillings. The interaction value is $J/t=0.8$, for
  which the SDW phase exists until $\langle
  n\rangle\approx 0.6$.
\label{fourier}}
\end{figure}
The observed modulations of the spin polarization and the electron
density are not independent of model parameters.
In order to obtain information about the wavelength of the modulations,
we have performed a Fourier
transformation of the obtained DMFT solutions. For this purpose, we
have used $(60\times 10)$-clusters for our calculations.  
The Fourier transformation for the occupation and the polarization read
\begin{eqnarray*}
\tilde n(k_x,k_y)&=&\frac{1}{N}\sum_{x,y}e^{ik_xx+ik_yy}(n_{\uparrow
  (x,y)}+n_{\downarrow (x,y)}-n_{aver})\\
\tilde m(k_x,k_y)&=&\frac{1}{N}\sum_{x,y}e^{ik_xx+ik_yy}(n_{\uparrow
  (x,y)}-n_{\downarrow (x,y)}),\\
\end{eqnarray*}
where we have subtracted the average filling of the lattice,
$n_{aver}=\frac{1}{N}\sum_{x,y}(n_{\uparrow (x,y)}+n_{\downarrow
  (x,y)})$, and $N$ corresponds to the number of lattice sites. The
absolute values of these Fourier modes show distinct peaks at
certain momenta in the Brillouin zone. 
We show these momenta in Fig. \ref{fourier} for
$J/t=0.8$. For this interaction strength, the SDW phase exists until close to
quarter filling, where it changes into another magnetic
phase via a first order transition. The wavelength does not depend on
the coupling strength $J$ within the 
weak coupling region of the phase diagram.
If we neglect the parameter region of small hole doping, where the
wavelength of the SDW exceeds the 
cluster size, the largest Fourier component in the SDW and the CDW
can be fitted by a linear function in the number of holes. 
Exactly at half filling, we find a N\'eel state with modulation
$\vec{k}=(\pi,\pi)$ without a CDW. Away from half filling this N\'eel
state changes into an SDW with modulation $\vec{k}=\pi(1-\langle h\rangle,1)$
accompanied by a CDW with modulation $\vec{k}=\pi(2\langle
h\rangle,0)$, where $\langle h\rangle=1-\langle n\rangle$ corresponds
to the number of holes in the lattice. 
For stronger interaction values,
the SDW phase vanishes upon doping 
into the paramagnetic heavy fermion state 
for small hole doping. The wavelength in this parameter region is
too long, as that we could 
accurately analyze the doping dependence of it.

\section{spectral function and Fermi surface}
As shown in Fig. \ref{phase_diagram}, we have divided the SDW phase
into two regions; one with large and one with small Fermi surface. We
next want to 
show momentum-resolved spectral functions and the corresponding
Fermi surfaces which justify this distinction.

\begin{figure*}
\includegraphics[width=0.32\linewidth]{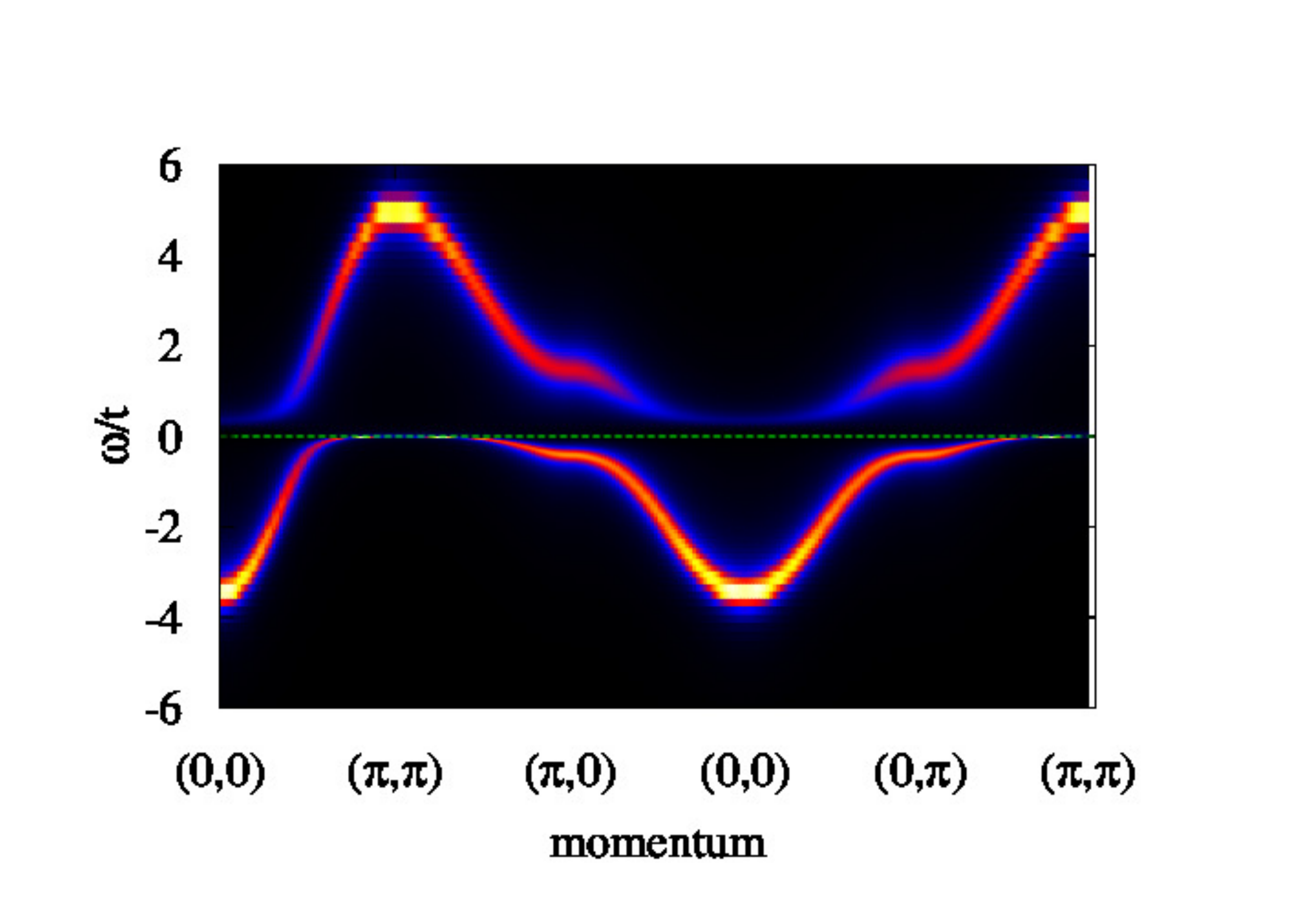}
\includegraphics[width=0.32\linewidth]{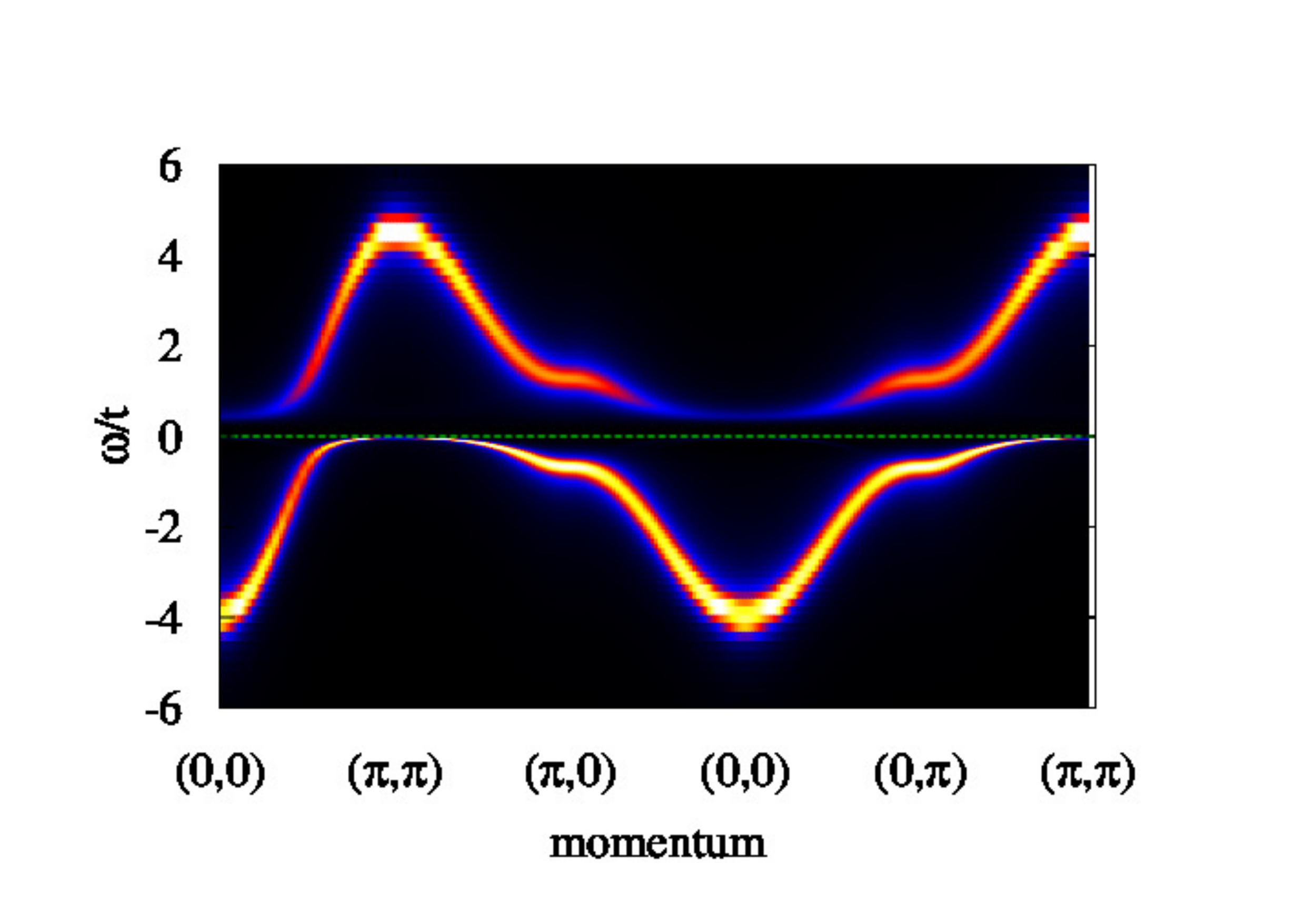}
\includegraphics[width=0.32\linewidth]{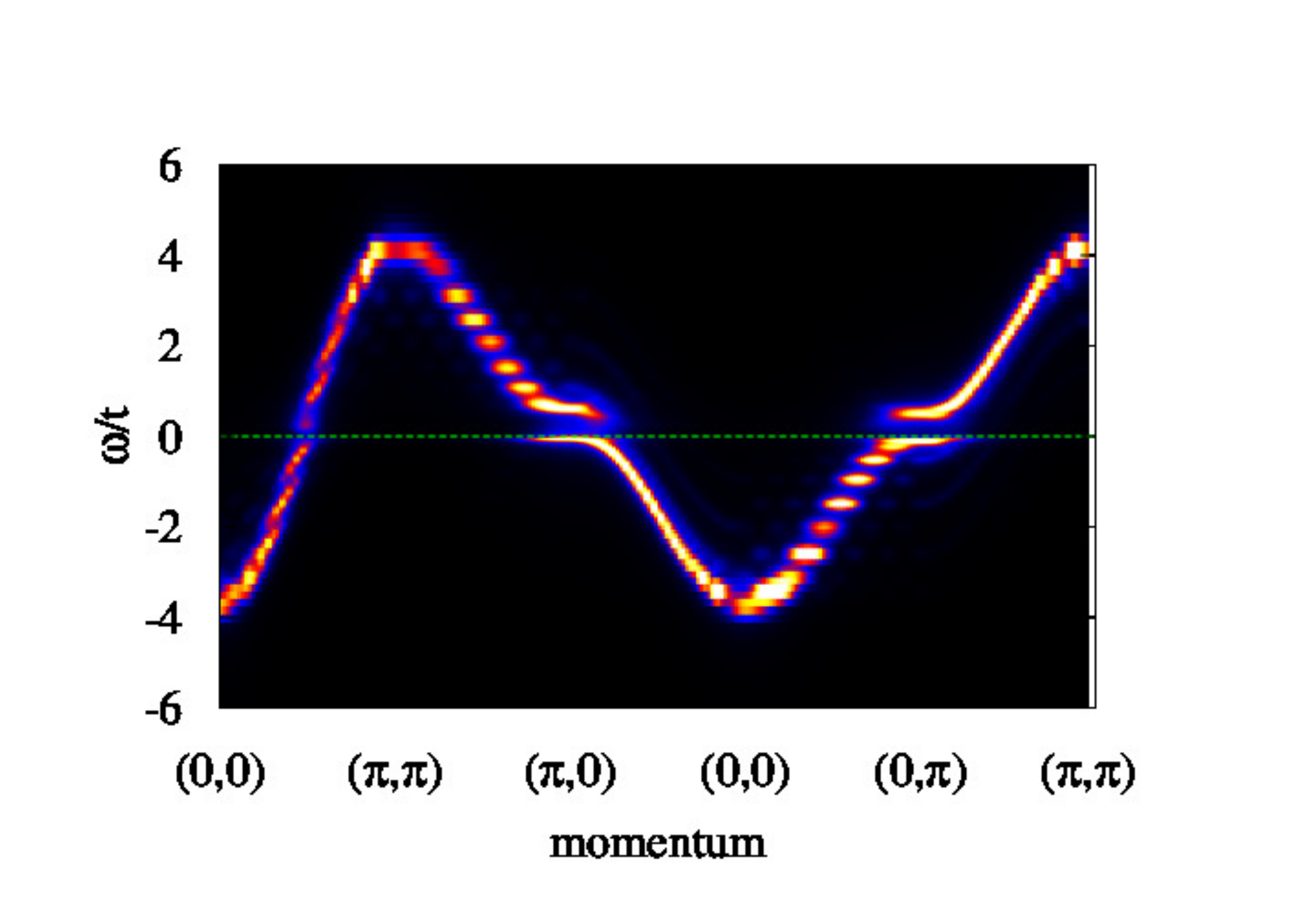}\\
\includegraphics[width=0.32\linewidth]{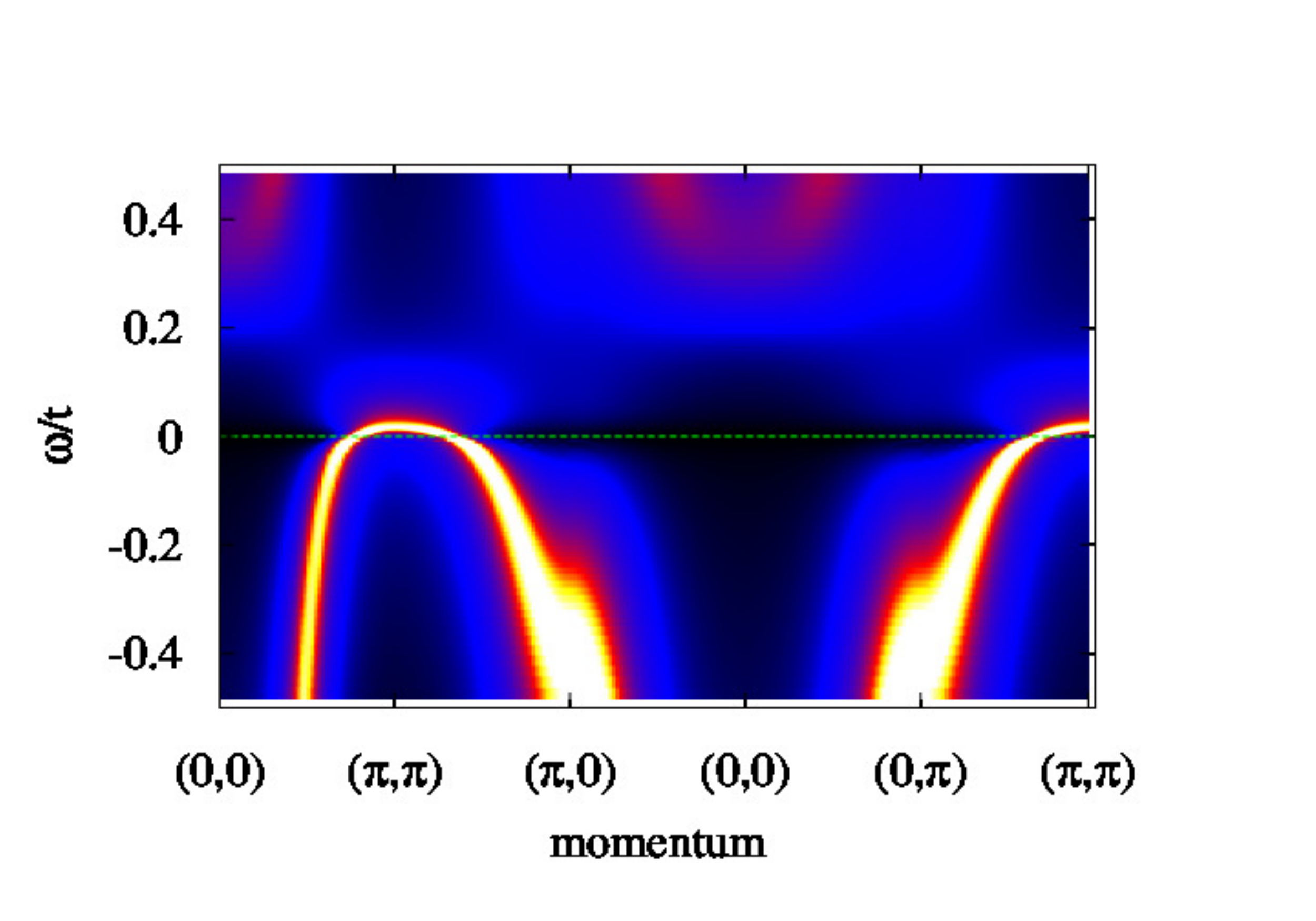}
\includegraphics[width=0.32\linewidth]{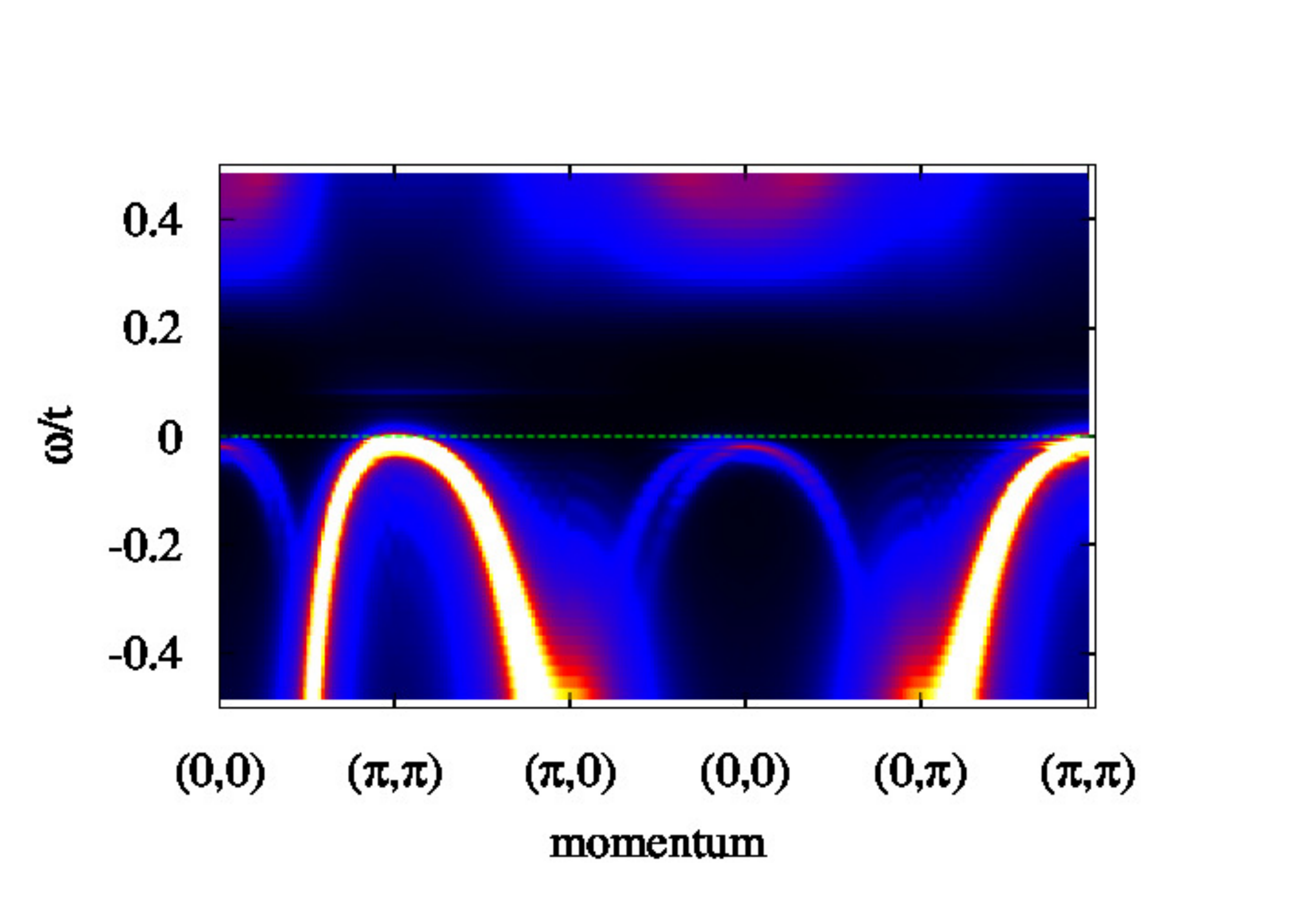}
\includegraphics[width=0.32\linewidth]{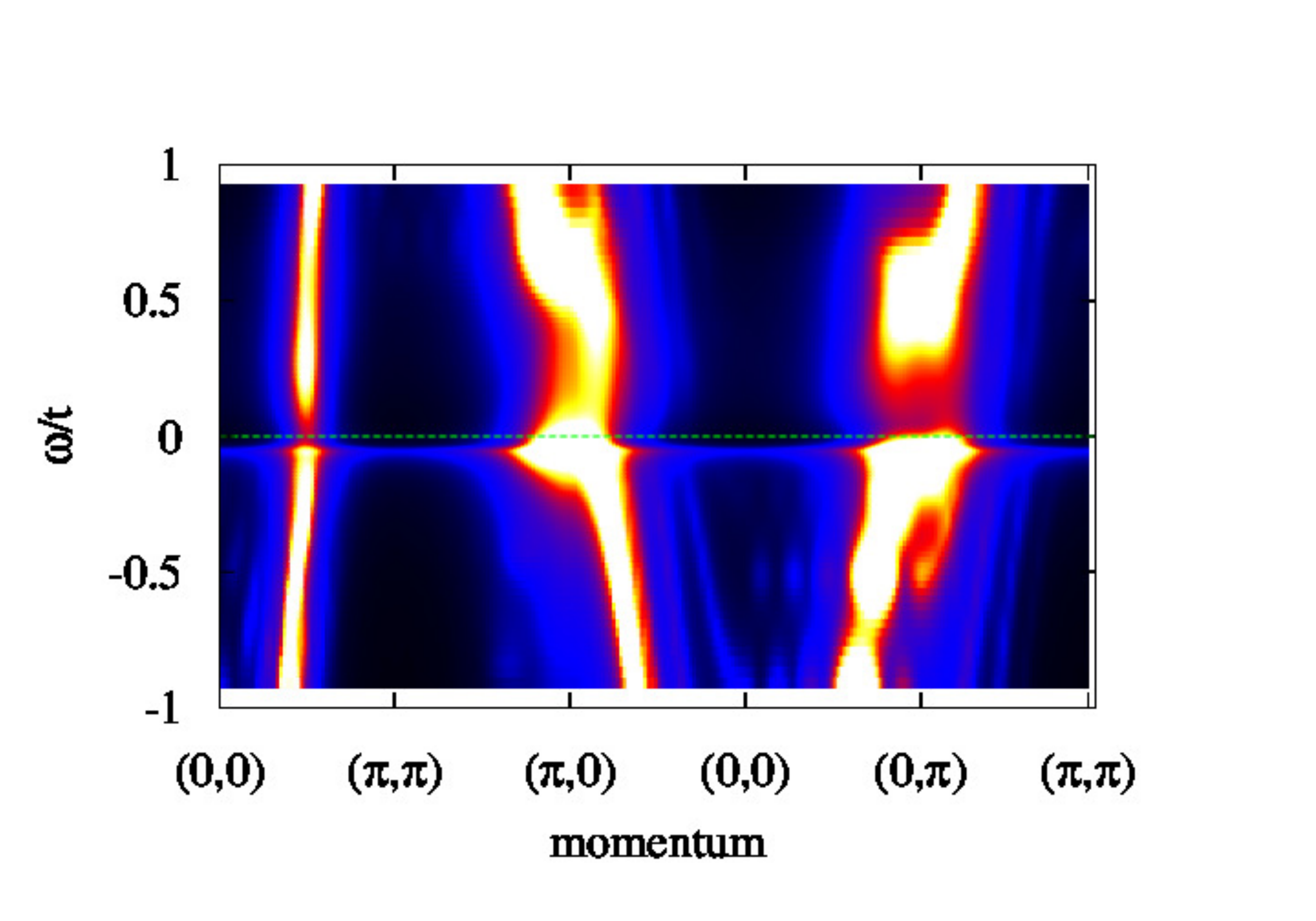}
\caption{Spectral functions for the paramagnetic state (left), the
  strong-coupling SDW (middle), and the weak-coupling SDW state
  (right). The lower panel is a magnification of the upper panel
  around the Fermi energy.
\label{specs}}
\end{figure*}

Figure \ref{specs} shows spectral functions for the paramagnetic
phase (left), the strong-coupling SDW phase (middle), and the
weak-coupling SDW phase (right). The lower panels correspond to 
magnifications around the Fermi energy of the upper panels. The green
dashed line marks the Fermi energy.
In the weak-coupling SDW phase (right panel in Fig. \ref{specs}), we
only find
slight modifications from the non-interacting energy momentum
dispersion. These modifications mainly occur close to the Fermi energy
around $(\pi,0)$, $(0,\pi)$, and $(\pi/2,\pi/2)$, where we observe
a suppression of the spectral weight. 
Compared to the non-interacting spectrum, additional weak bands occur
around $(0,0)$ and $(\pi,\pi)$ below the Fermi energy. Especially the
band which is close to the Fermi energy at $(\pi,\pi)$ is a remnant of
the heavy fermion band, which however lies below the Fermi
energy. The main contribution to the spectral weight at the 
Fermi energy originates from the non-interacting dispersion at
$(\pi,0)$, $(0,\pi)$, and $(\pi/2,\pi/2)$.

The spectral functions of the paramagnetic metallic phase (left panel
in Fig. \ref{specs}) and the strong-coupling SDW phase (middle
panel), differ substantially from the weak-coupling case. The bands, which
cross the Fermi 
energy in the non- and weakly-interacting case at  $(\pi,0)$ and
$(0,\pi)$, are absent at the Fermi energy in this phase.
While in the weakly-interacting SDW phase there is a band
which connects the low energy parts ($\omega/t=-4$) with the high
energy parts ($\omega/t=4$), in the paramagnetic and the
strong-coupling SDW state there is a gap above the Fermi
energy separating the low and the high energy parts of the spectrum.
The non-interacting bands, which approach the Fermi energy at
$(\pi,0)$, $(0,\pi)$, and $(\pi/2,\pi/2)$ are bent towards each other
and cross the Fermi energy close to $(\pi,\pi)$. The spectrum of the
paramagnetic state and the strong-coupling SDW state look thereby very
similar to each other. The gap above the Fermi energy seems to be more
pronounced in the SDW state. However, this is parameter dependent. Another
general feature in the dispersion of the SDW state is that
there is a band at $(0,0)$ close to the Fermi energy. This is a shadow
band originating from the ordered state.\cite{Bauer2007,Peters2014}

\begin{figure*}
\begin{center}
\includegraphics[width=0.195\linewidth]{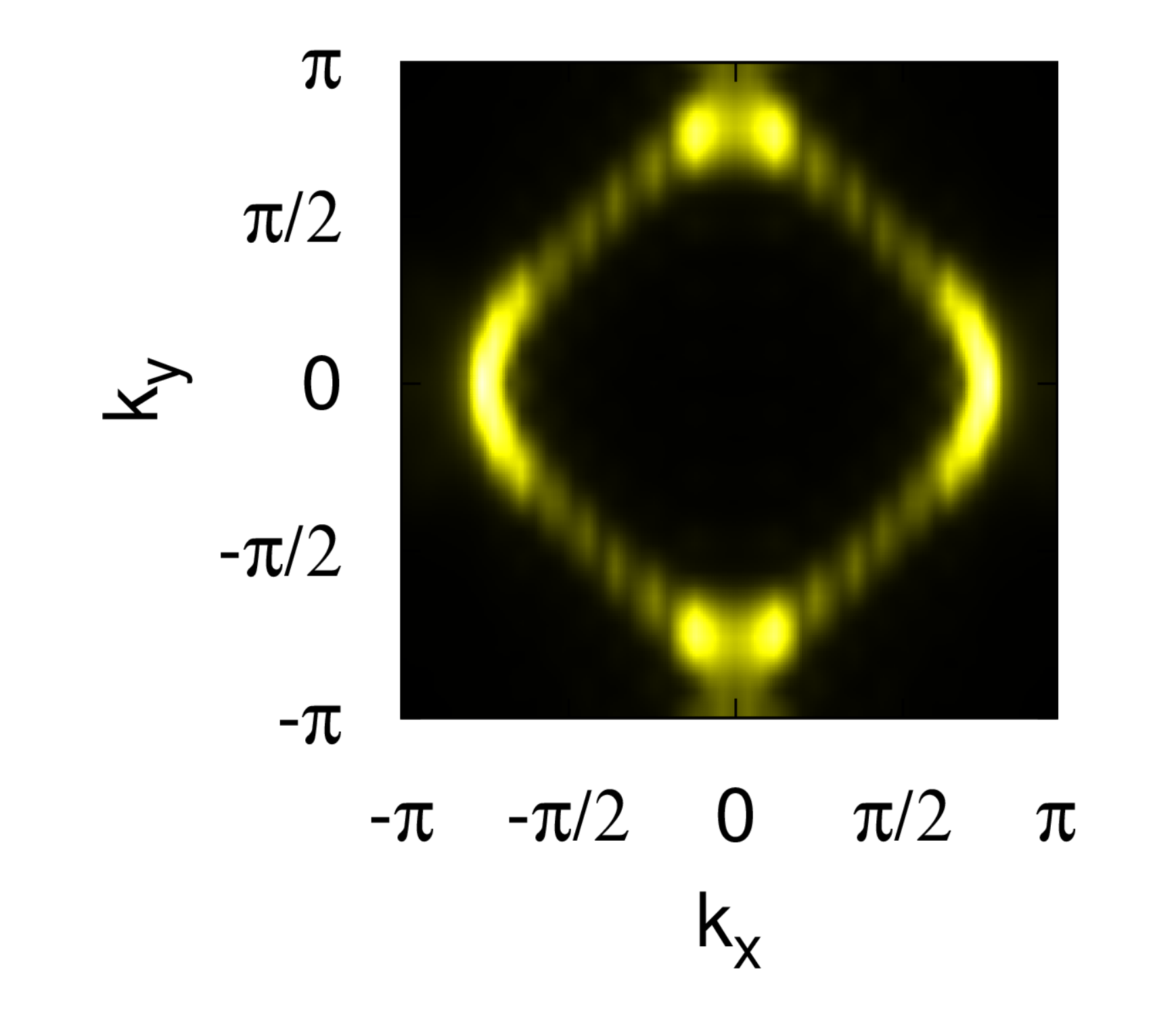}
\includegraphics[width=0.195\linewidth]{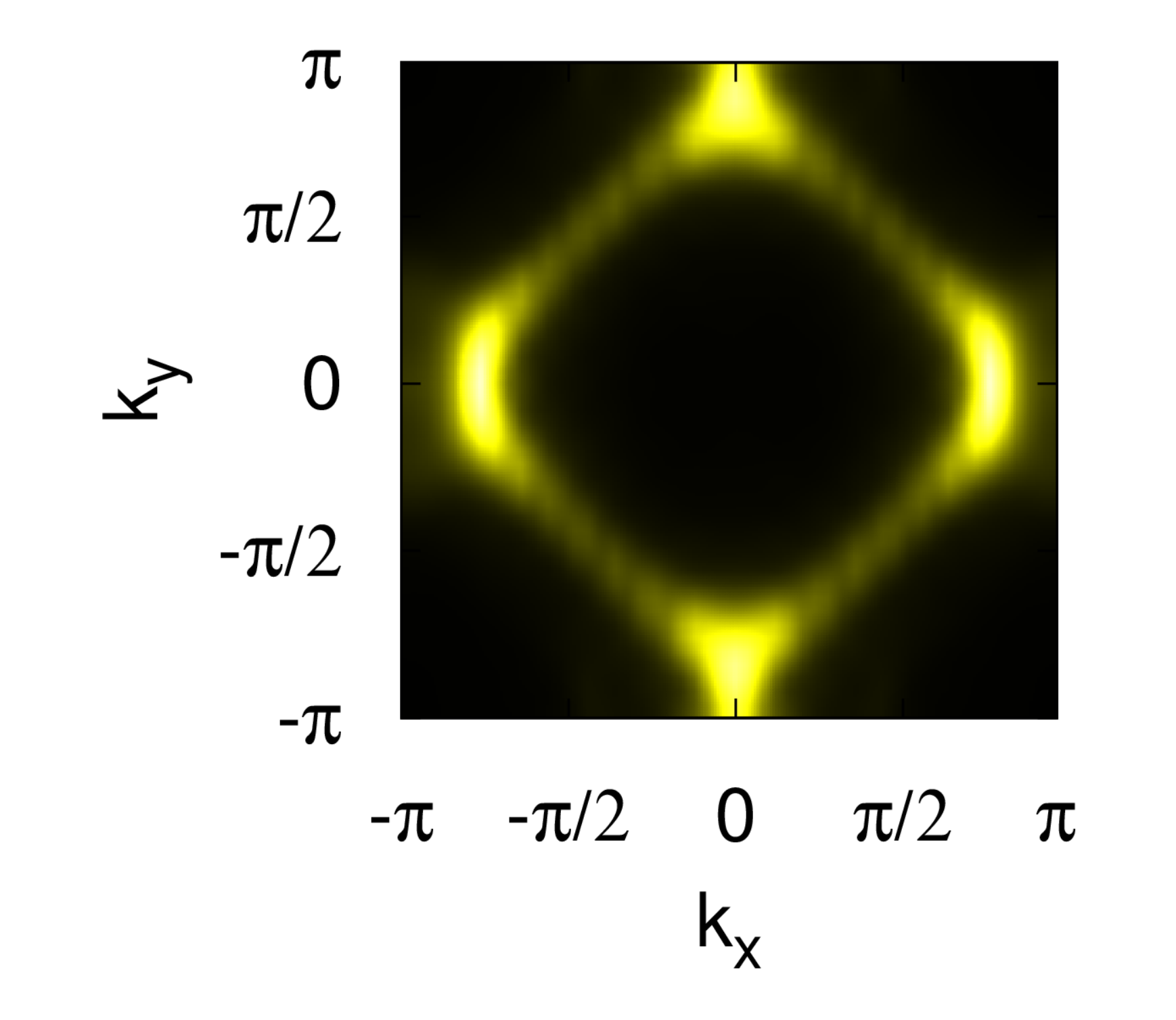}
\includegraphics[width=0.195\linewidth]{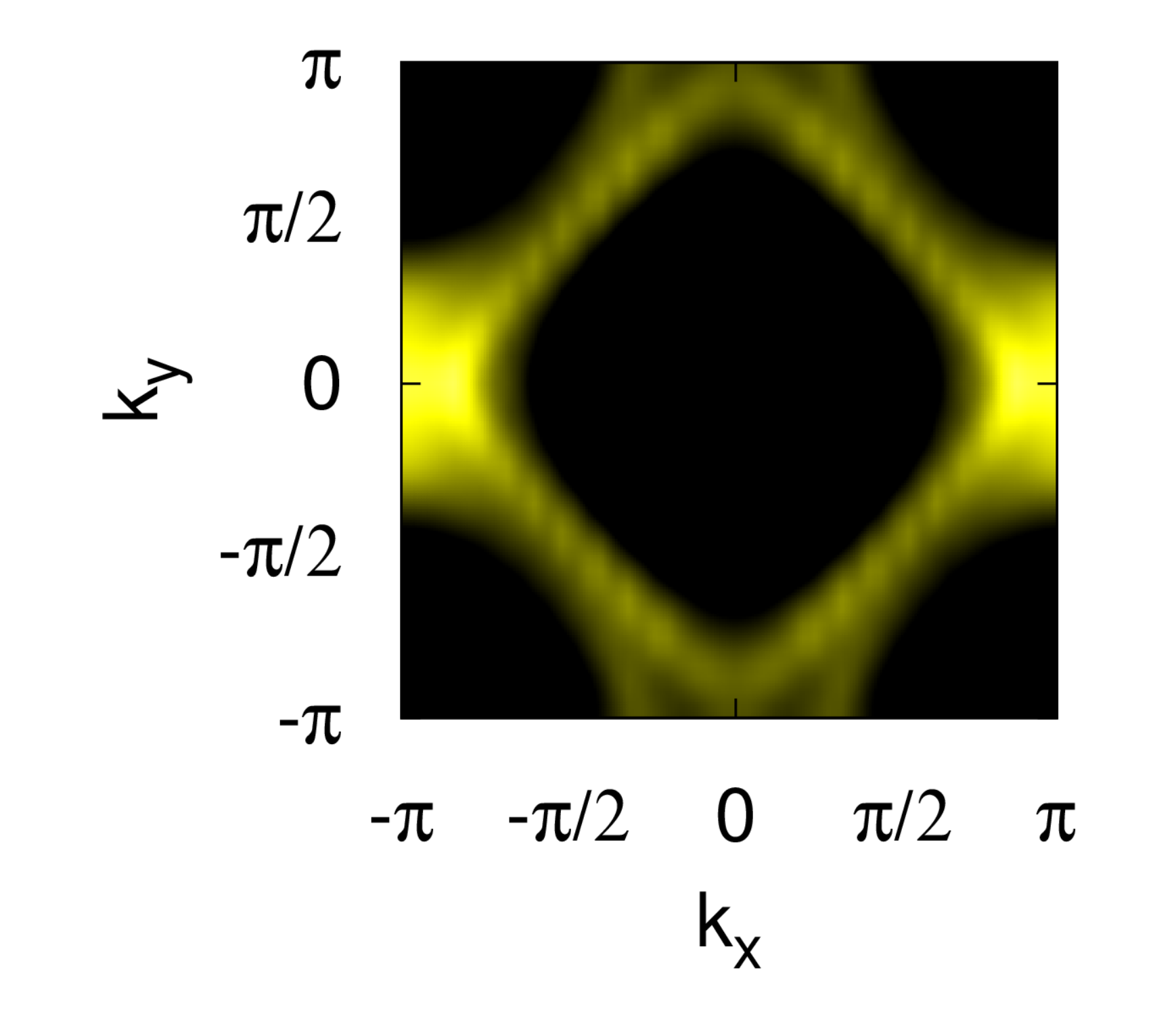}
\includegraphics[width=0.195\linewidth]{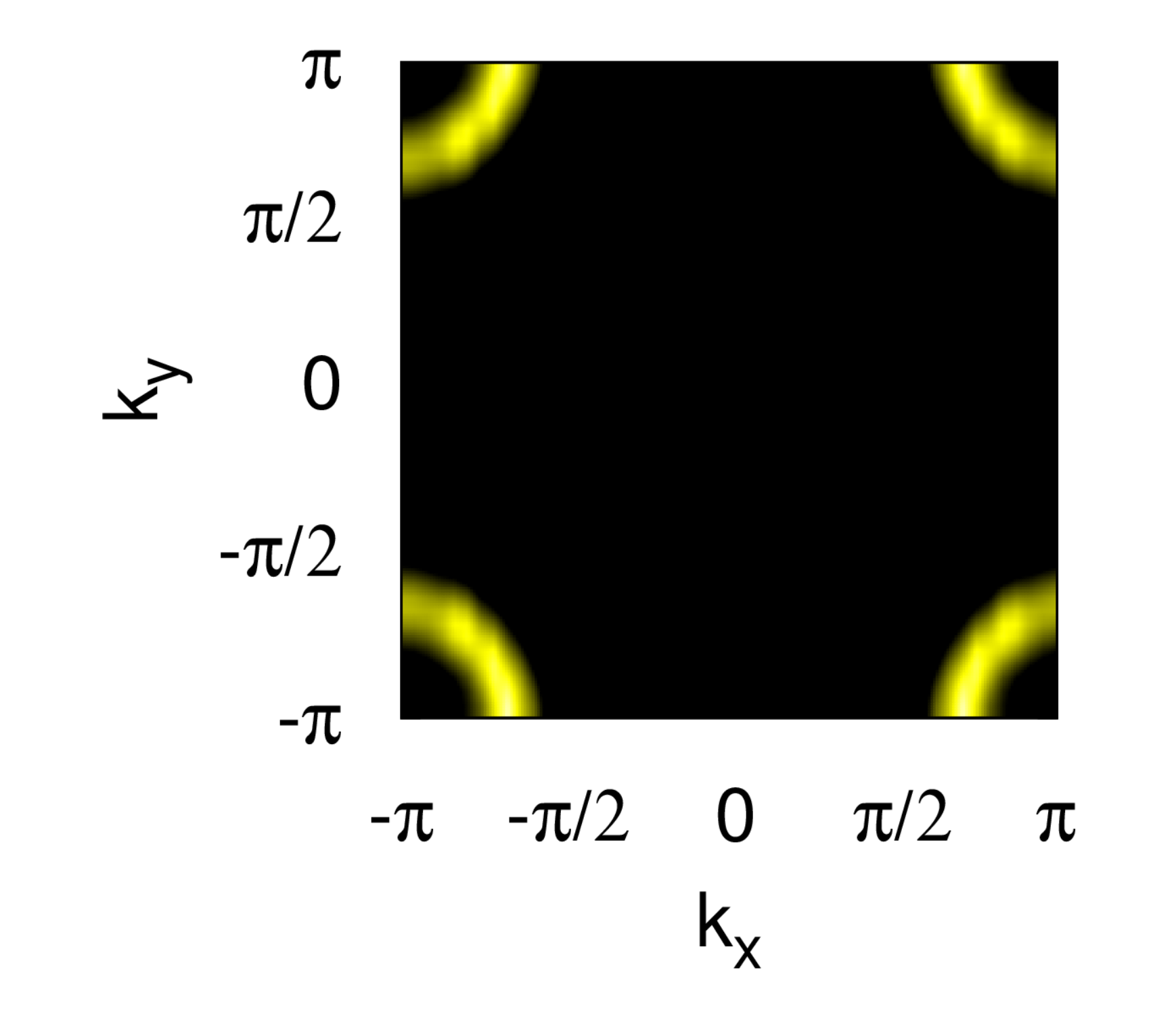}
\includegraphics[width=0.195\linewidth]{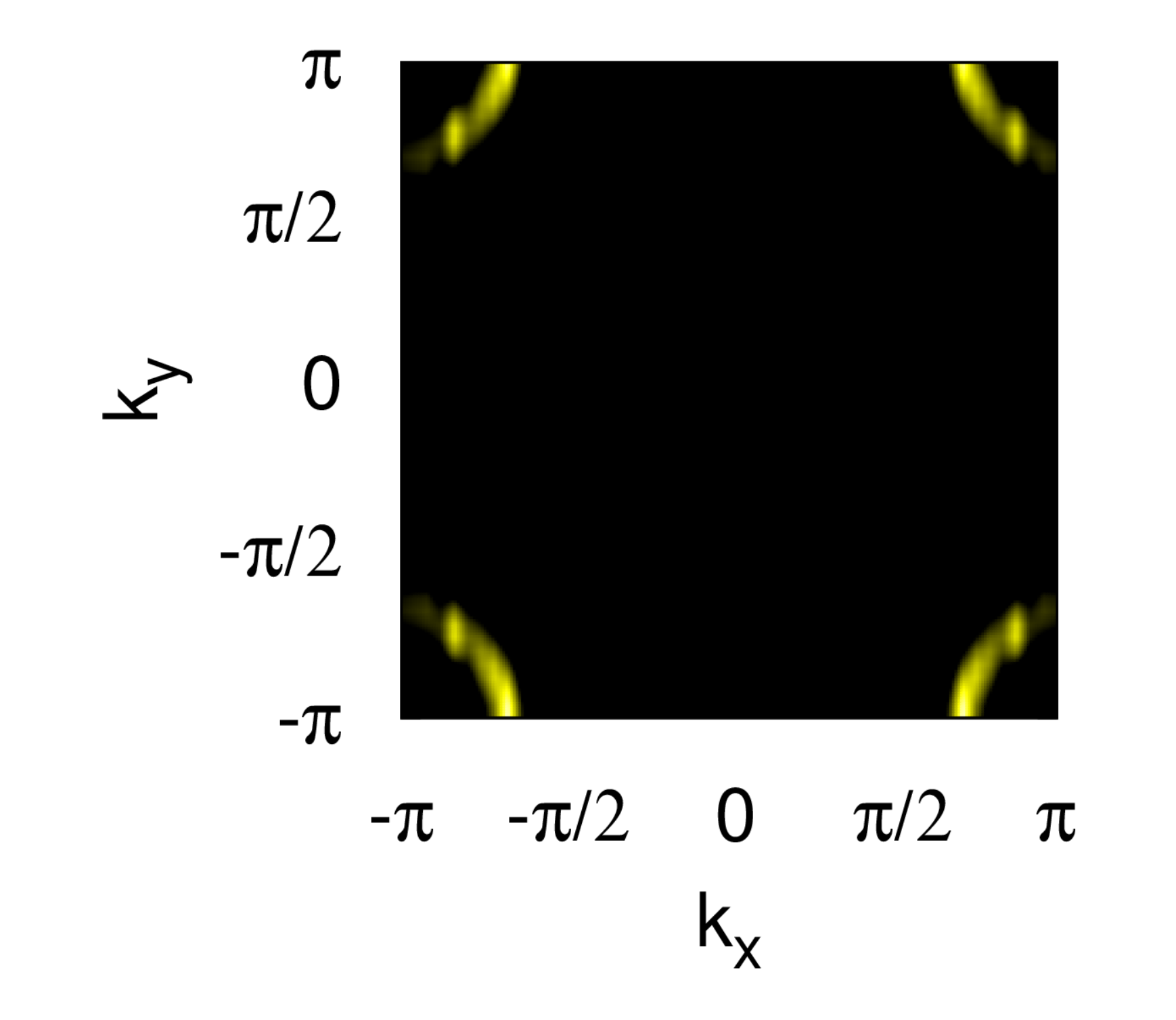}\\
\includegraphics[width=0.195\linewidth]{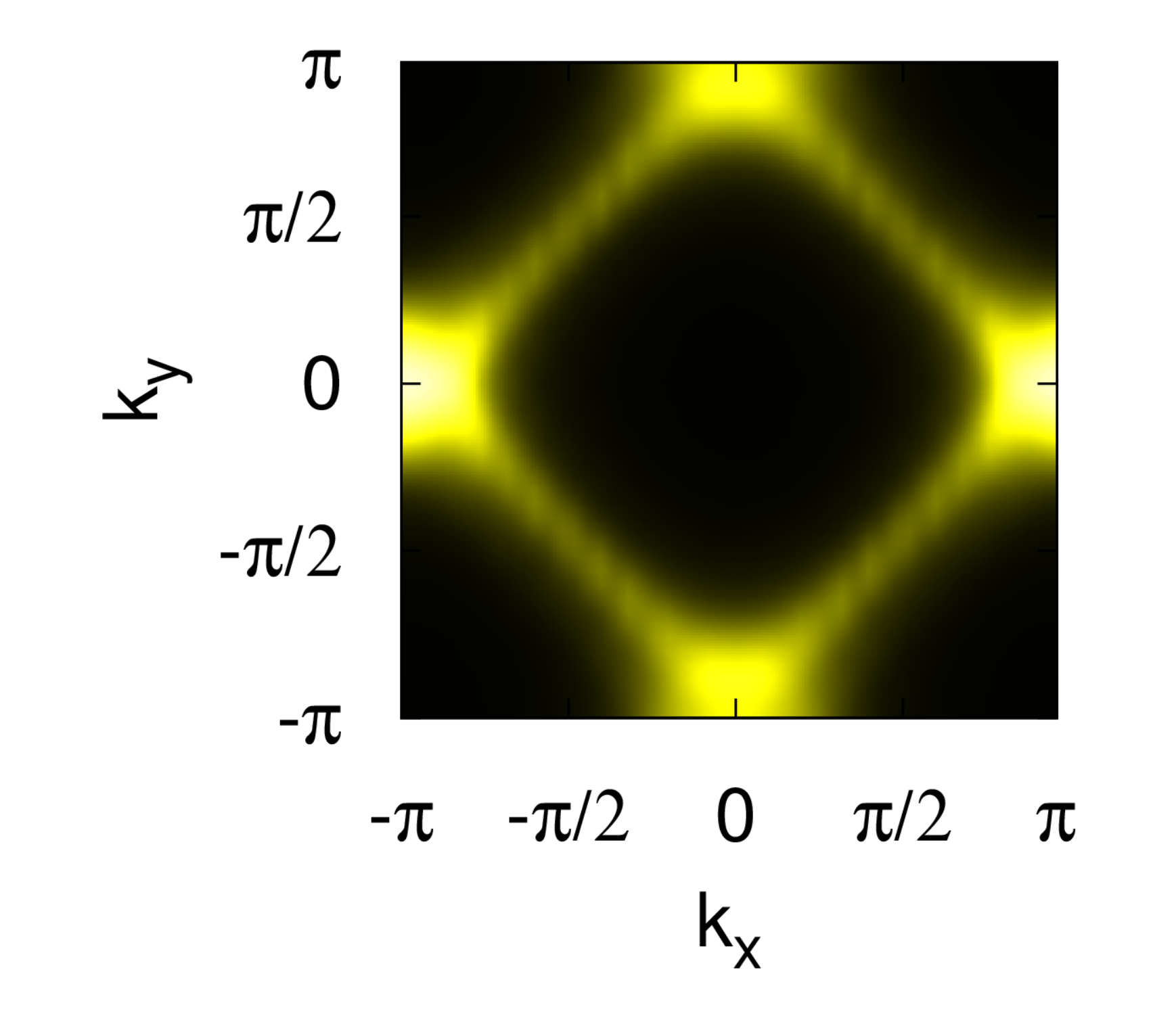}
\includegraphics[width=0.195\linewidth]{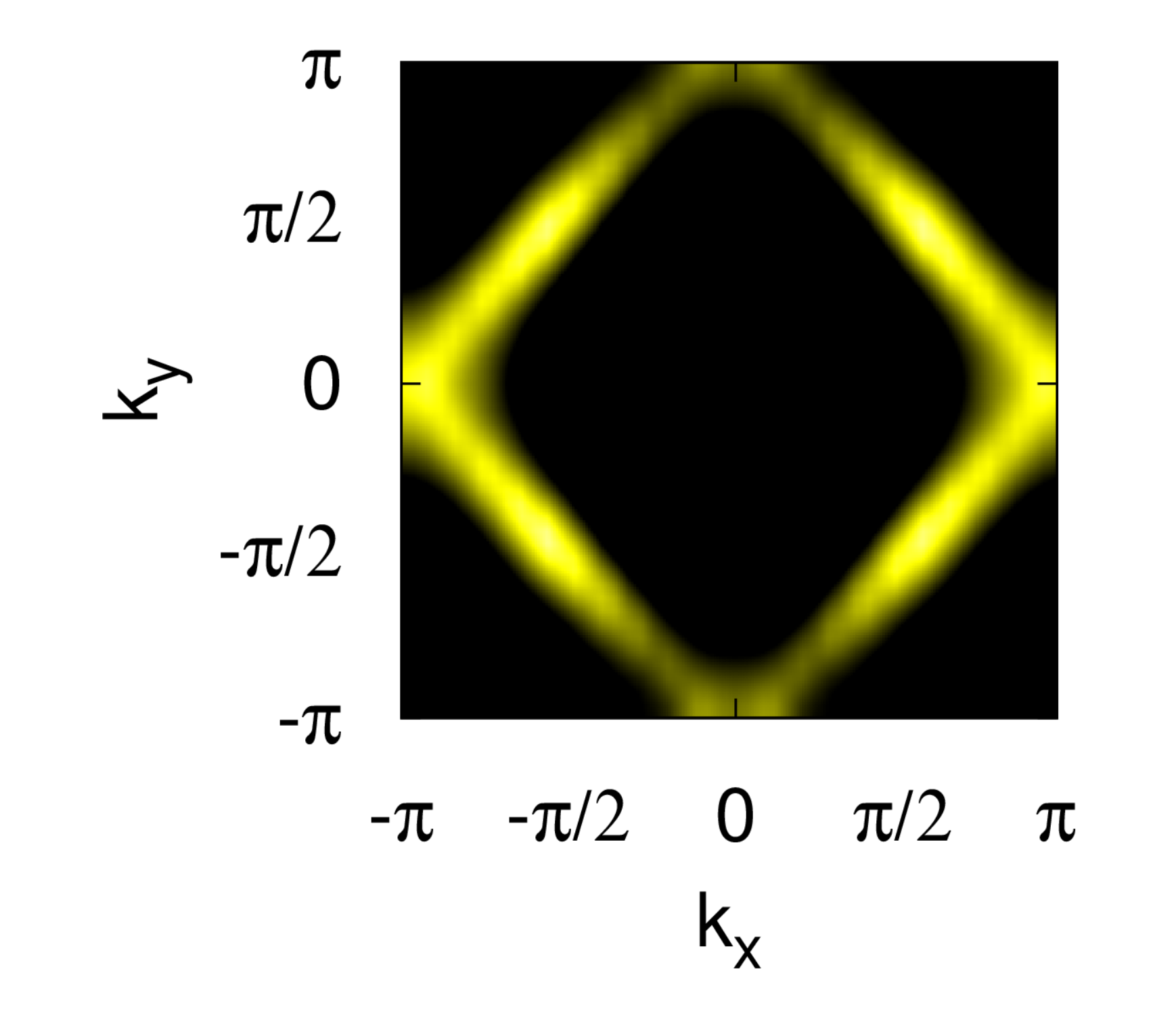}
\includegraphics[width=0.195\linewidth]{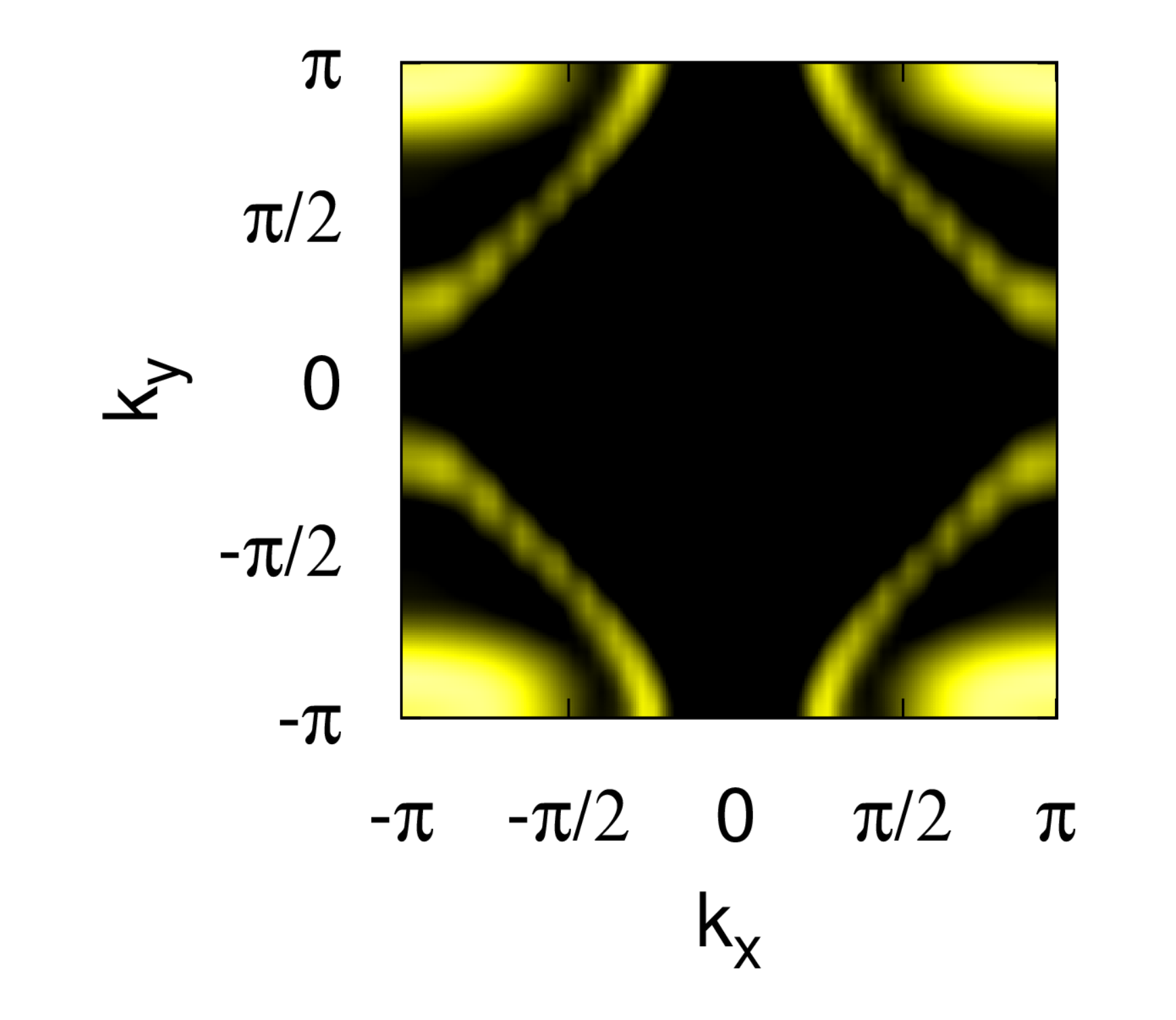}
\includegraphics[width=0.195\linewidth]{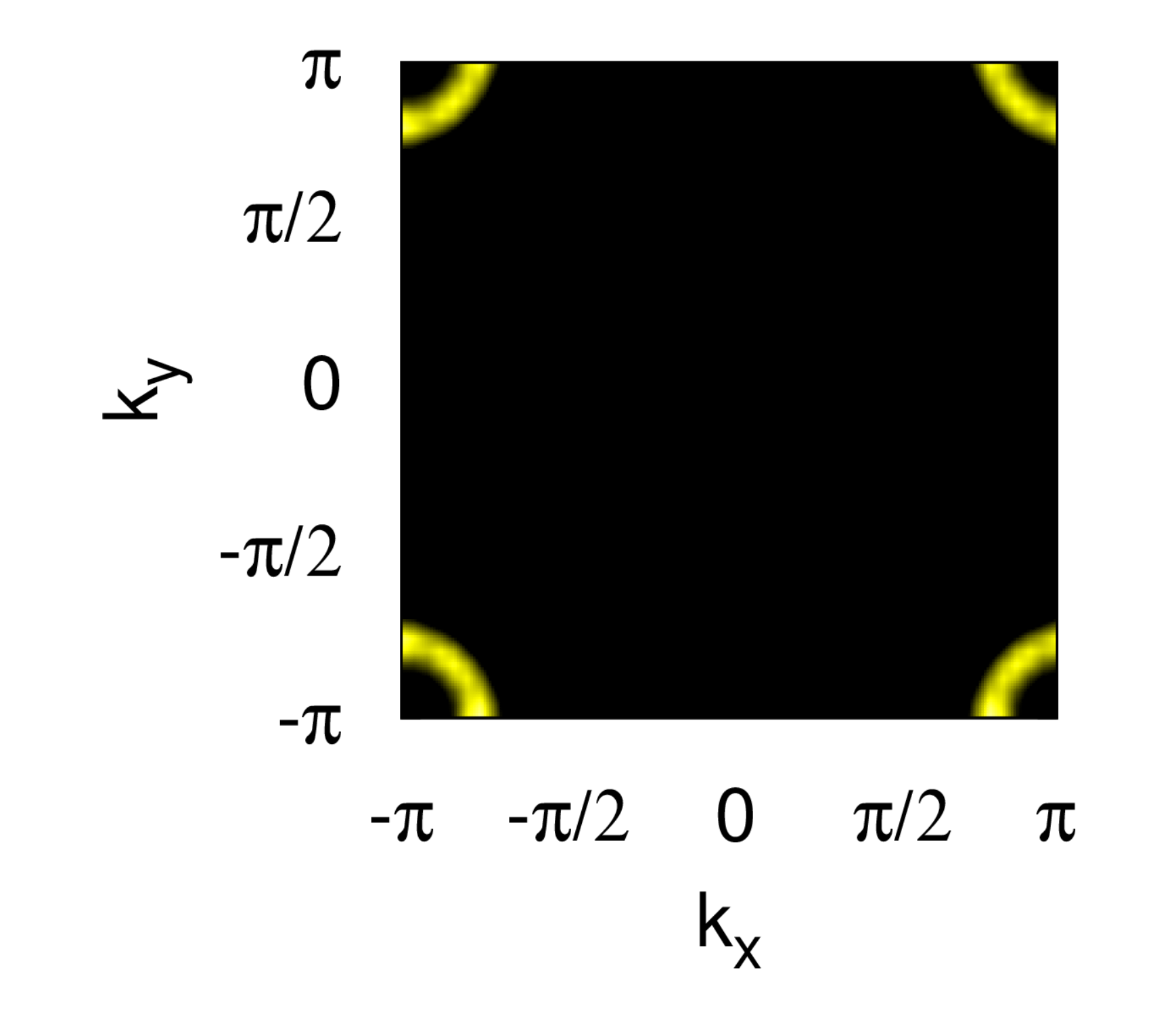}\\
\includegraphics[width=0.195\linewidth]{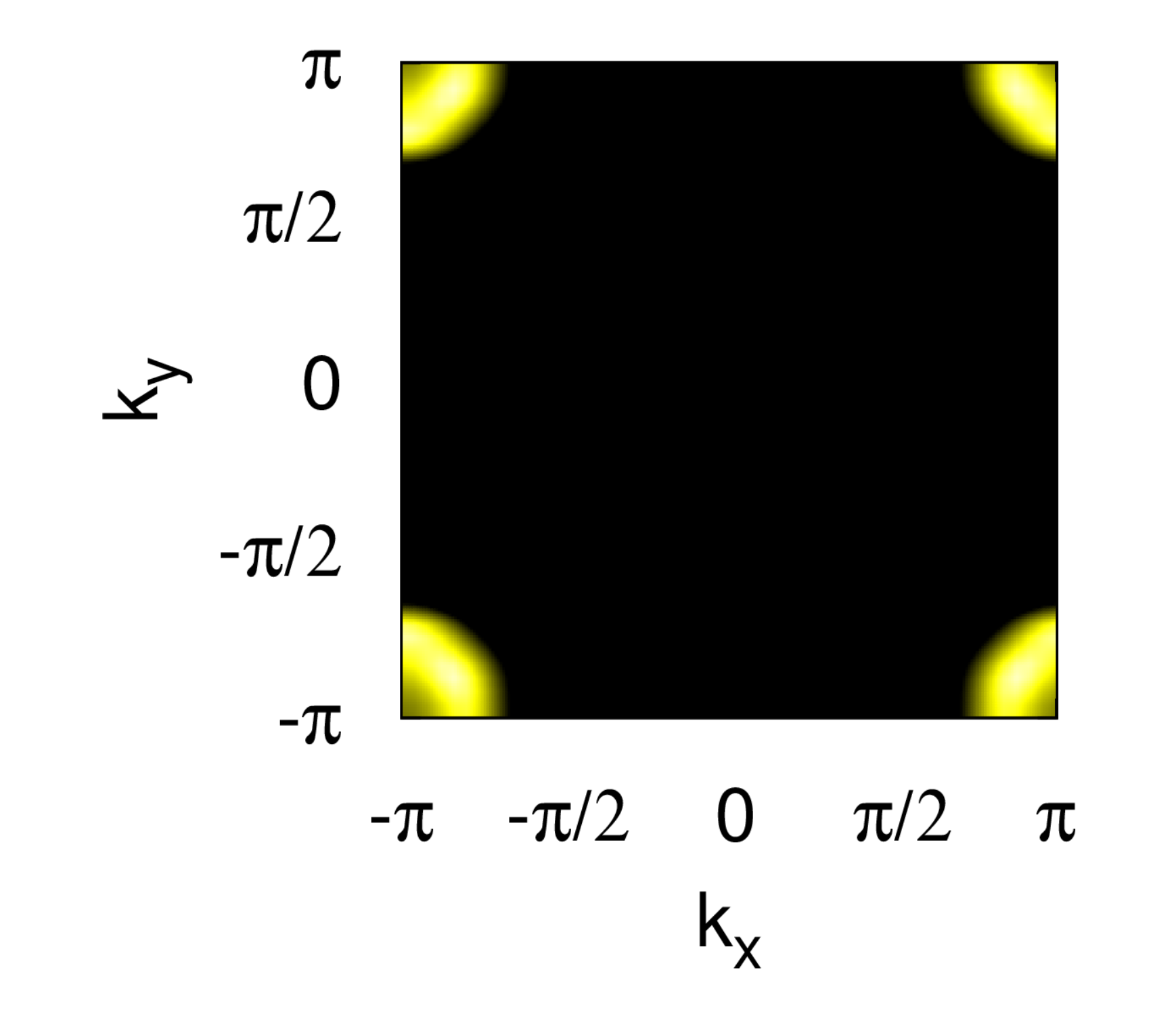}
\includegraphics[width=0.195\linewidth]{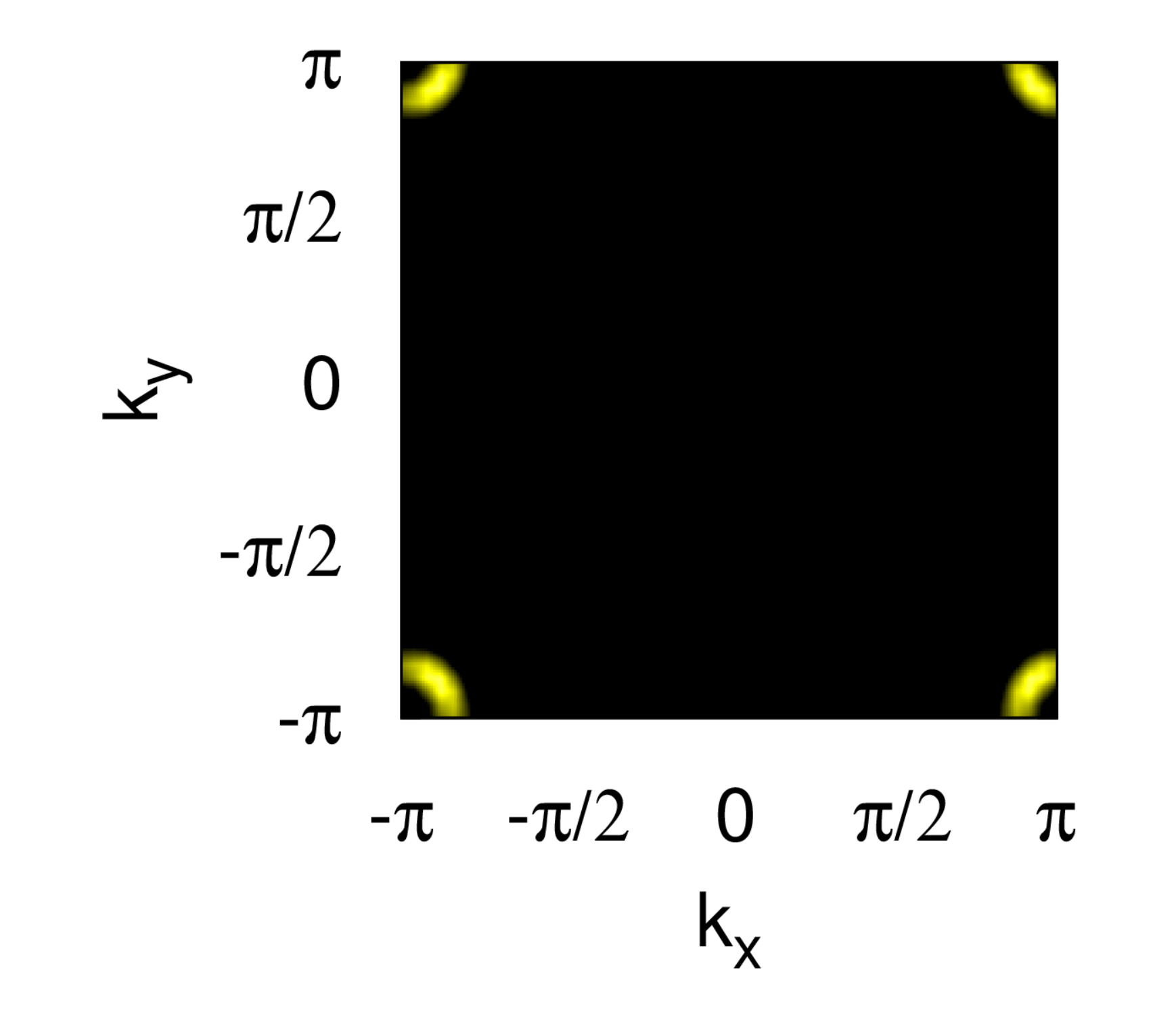}
\includegraphics[width=0.195\linewidth]{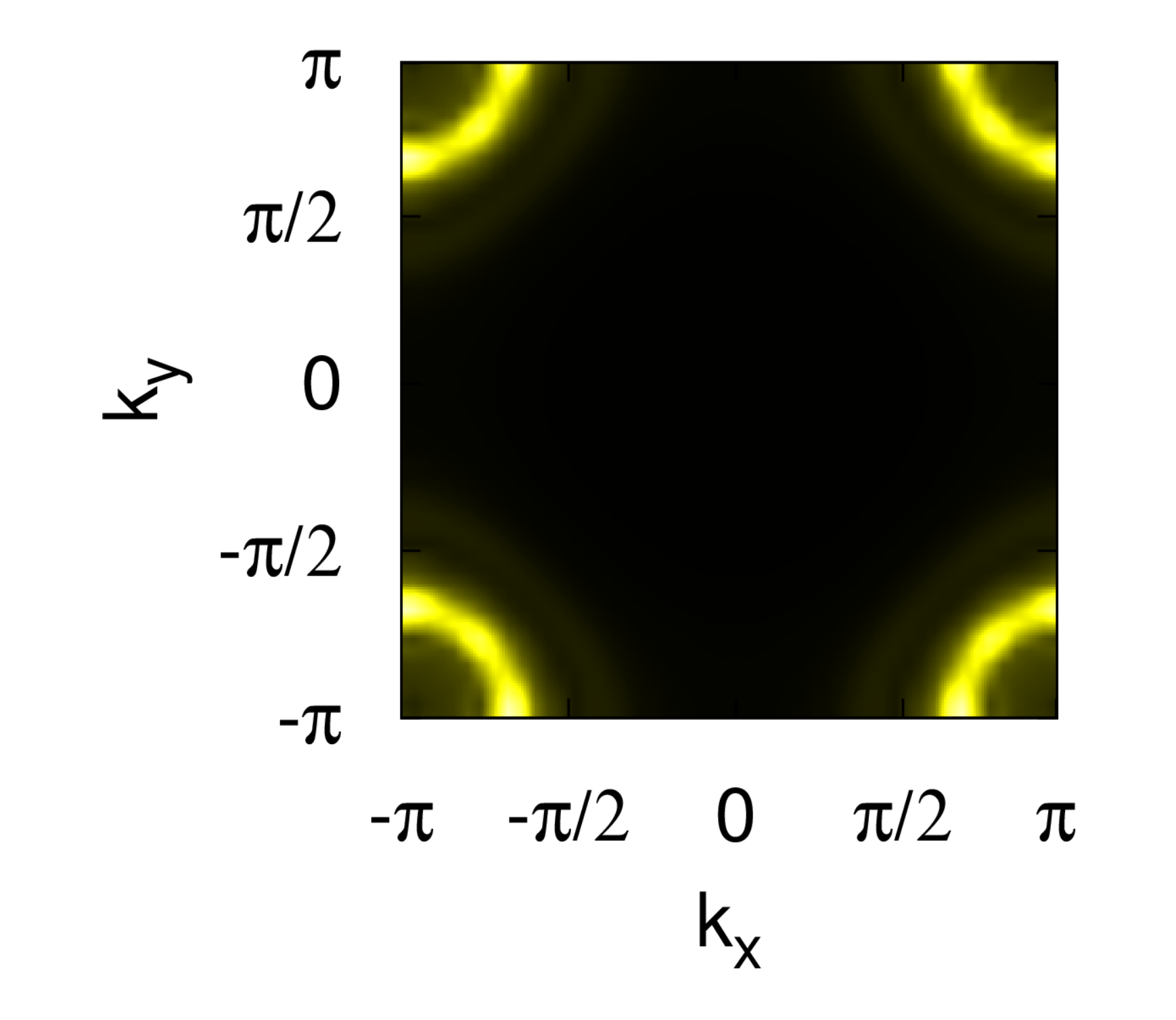}
\end{center}
\caption{Calculated Fermi surfaces. Top panels:
  Fermi surfaces across the transition between the weak-coupling SDW and
the paramagnetic state at conduction band filling  $\langle
  n\rangle\approx0.8$ and (from left to right)
  $J/t=(0.6;0.8;1.0;1.2;1.4)$. The transition takes place at $J=t$. 
Middle panels: Fermi surfaces
  inside the SDW phase for (from left to right)
  $J/t=(0.8;1.0;1.2;1.4)$ and $\langle n\rangle\approx 0.9$. The Fermi
  surface topology changes at $J/t=1.2$ (second from right). Bottom
  panels: Fermi surfaces for different dopings across
 the phase transition between the strong-coupling SDW and the
  paramagnetic state at $J/t=1.8$ for (from left to right) $\langle
  n\rangle=(0.98;0.93;0.85)$. The magnetic order vanishes in the
  middle panel.
\label{Fermi}}
\end{figure*}
From the spectral functions shown in Fig. \ref{specs}, we can
see that the Fermi surface changes within the SDW phase. For a better
understanding, we have
compiled the calculated Fermi surfaces for a few parameters in
Fig. \ref{Fermi}. Different rows show different transitions, which can
be observed in the Kondo lattice model. The upper panels show the
transition between the weak-coupling SDW phase and the paramagnetic phase at
$\langle n\rangle\approx0.8$ for $J/t=(0.6;0.8;1.0;1.2;1.4)$
(left-right). One can clearly observe a change in the Fermi surface
topology at $J/t=1$. While in the left two panels the Fermi surface is
small including only conduction electrons, the Fermi surface in the
right two panels is large, which includes conduction electrons and
localized moments. A similar change in the Fermi surface topology can
be observed when increasing the interaction strength within the SDW
phase (middle panels of Fig. \ref{Fermi}).  The Fermi surface clearly
changes in the second panel from right for $J/t=1.2$ and $\langle
n\rangle\approx0.9$. Finally, the lower panels show the transition
between the strong-coupling SDW and the paramagnetic phase for
different conduction band fillings and $J/t=1.8$. The magnetic order
vanishes in the middle panel for $\langle
n\rangle\approx0.93$. However, there is no abrupt change in the Fermi
surface for strong coupling when the magnetic order vanishes.
Summarizing these results, we find that the SDW phase away from half
filling is divided into a phase with small Fermi surface at weak coupling and
a phase with large Fermi surface at strong coupling. We thus find a Lifshitz
transition within the SDW phase, where the Fermi surface topology
changes.

Although the Fermi surface vanishes exactly at half filling due to the
insulating nature of the system, a similar
change in the energy-momentum dispersion can be observed. At
weak coupling, the non-interacting bands approaching the 
Fermi energy at $(\pi,0)$, $(0,\pi)$, and $(\pi/2,\pi/2)$ are gapped
out at the Fermi energy. Bands at $(\pi,\pi)$ and $(0,0)$ do not
exist or are very weak. At approximately $J/t\approx 1.4$, this dispersion
changes. The bands approaching the Fermi energy are bent towards each
other. The spectral weight at $(\pi,0)$ and $(0,\pi)$ close to the
Fermi energy is shifted to $(\pi,\pi)$. Although none of these bands
crosses the Fermi energy, we nevertheless observe a similar change in the
energy-momentum dispersion at half filling as for the doped system.

\section{phase transitions}
As we have shown above, the Fermi surface changes within the SDW
phase. We will show now that the phase transition between the SDW
phase and the paramagnetic phase is also influenced by this change
inside the SDW phase.
\begin{figure}[tb]
\begin{center}
\includegraphics[width=\linewidth]{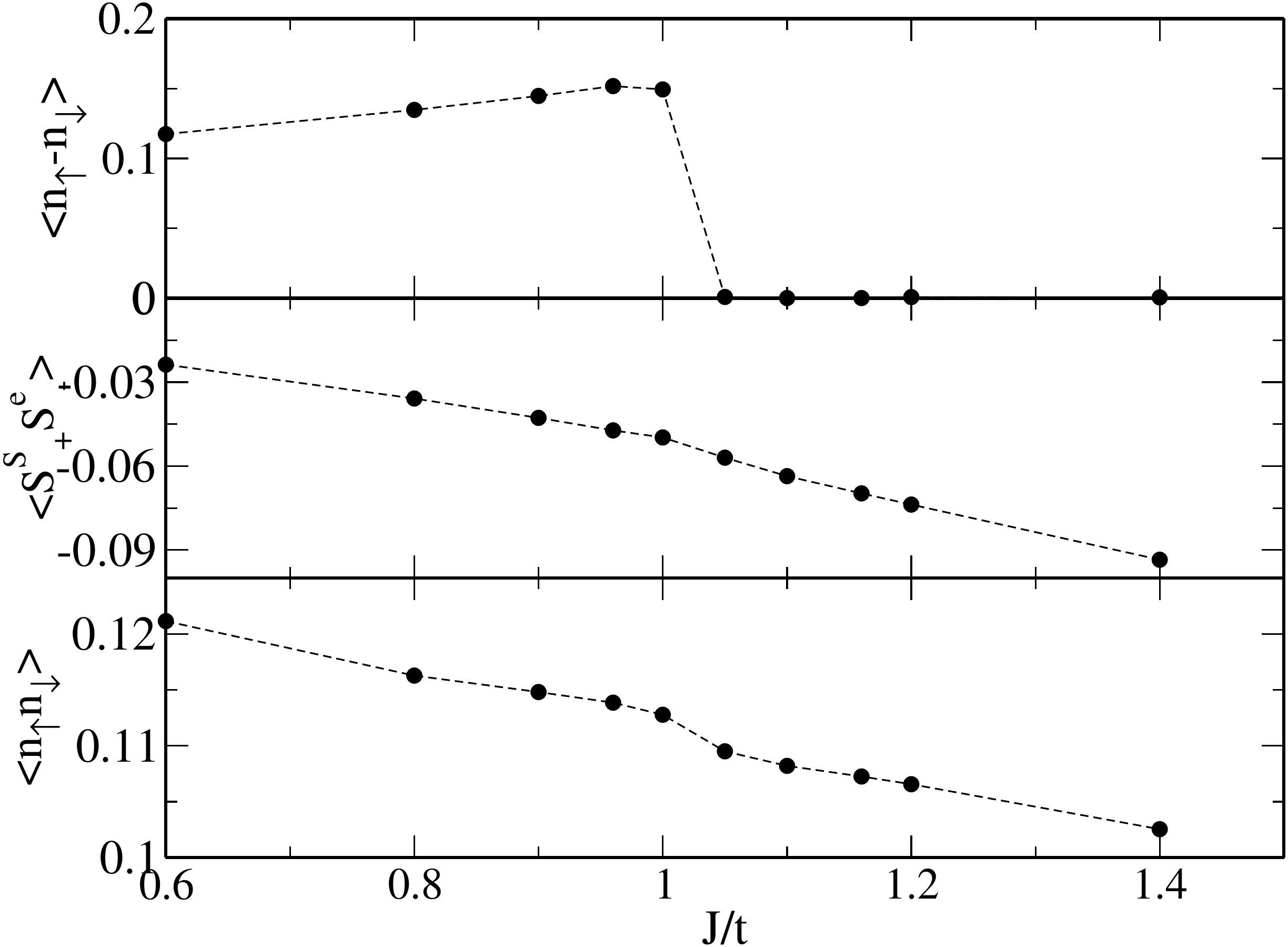}
\end{center}
\caption{First order phase transition between the 
  SDW phase and the paramagnetic phase at weak coupling and $\langle
  n\rangle\approx 0.8$. The spin polarization as well as the charge
  density order vanish abruptly at $J/t\approx 1$.  The figure shows
 the electron polarization (top), the spin-spin-correlation, $\langle
 S_+^cS_-^S\rangle $, 
 between conduction electrons and localized moments (middle),
 and the double occupancy of the conduction electrons (bottom). 
\label{trans_weak}}
\end{figure}
In Fig. \ref{trans_weak}, we analyze the transition between the SDW
state and the paramagnetic state at weak coupling and conduction band
filling $\langle n\rangle\approx 0.8$. This represents an example for
the transition between the small Fermi-surface SDW and the large
Fermi-surface paramagnetic state. We show in Fig. \ref{trans_weak} the 
polarization of the conduction electrons (top panel), the spin-spin
correlation between localized spins and conduction electrons (middle
panel), and the double occupancy of the conduction electrons (bottom
panel). We only show the expectation values corresponding to the
maximally polarized regions of the SDW.
As can be clearly seen, the electron polarization vanishes abruptly
around $J/t=1$, which signifies a first order transition between the
SDW phase and the paramagnetic phase. We note that within the SDW
phase also the electrons in the low electron-density regions are polarized
within this parameter region and that also this polarization jumps
abruptly to zero at the phase transition.
Exactly at this interaction strength, the Fermi surface changes
from small to large. 
We also observe small discontinuities in the spin-spin correlation and
the double occupancy at the phase transition. However, both
discontinuities are not very significant. We furthermore note that
although the Fermi surface becomes small, the
spin-spin correlation does not vanish at weak coupling.

\begin{figure}[tb]
\begin{center}
\includegraphics[width=\linewidth]{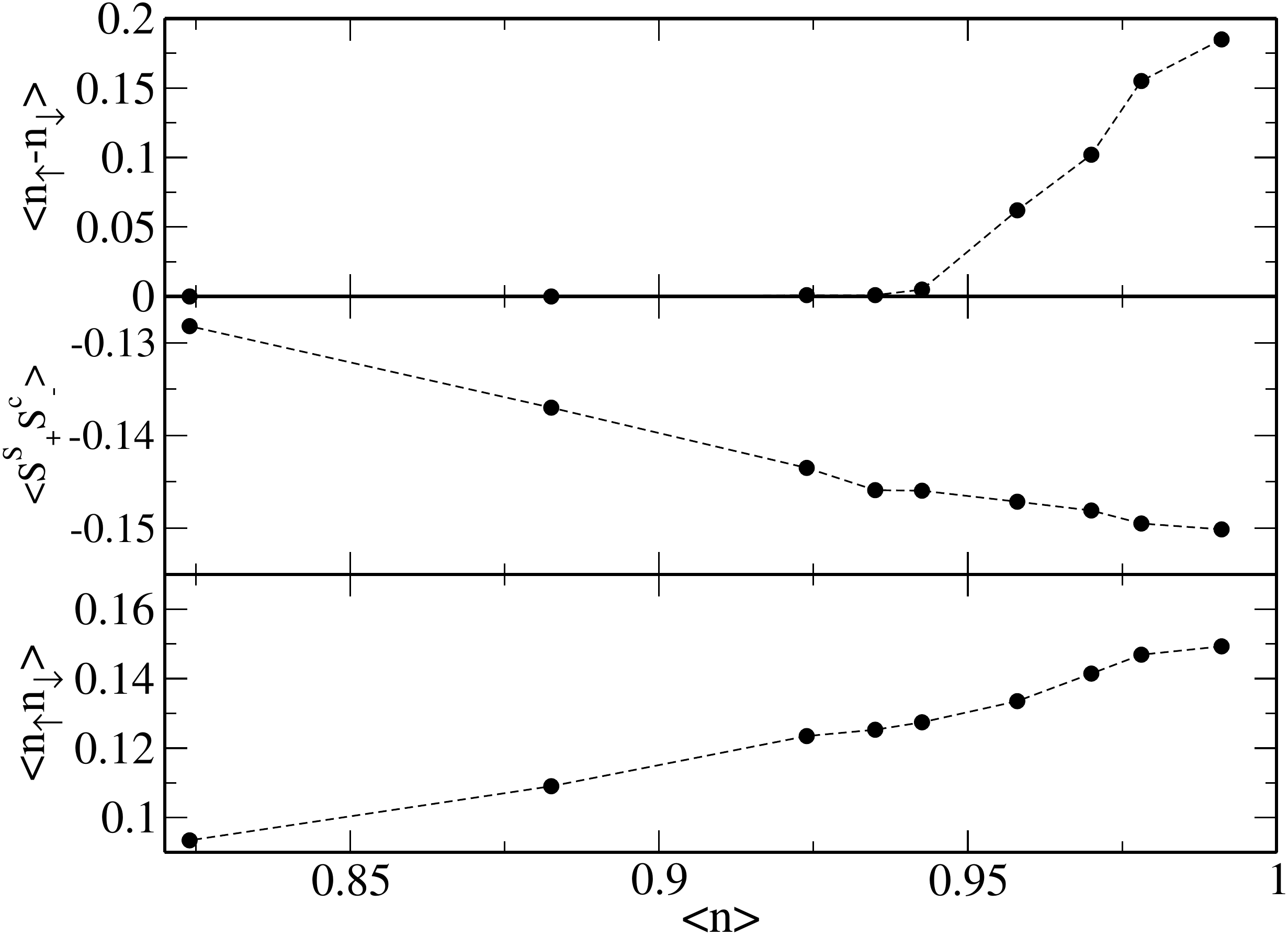}
\end{center}
\caption{Second order phase transition at $J/t=1.8$ for different
  conduction band
  fillings. The figure shows the same quantities as in Fig. \ref{trans_weak}.
\label{trans_strong}}
\end{figure}

The transition at strong coupling between the SDW
phase and the paramagnetic phase looks completely
different. We show the polarization of the conduction electrons (top panel), the spin-spin
correlation between localized spins and conduction electrons (middle
panel), and the double occupancy of the conduction electrons (bottom
panel) for conduction-band fillings across the phase transition at
$J/t=1.8$ in Fig. \ref{trans_strong}.
Contrary to the phase transition at weak coupling, the electron
polarization behaves smoothly upon doping at strong coupling. We observe the
well-known second order phase transition between the SDW phase and the
paramagnetic phase. 
Also
the spin-spin correlation and the double occupancy seem to be smooth across
the phase transition, although there might be small kinks.

\begin{figure}[tb]
\begin{center}
\includegraphics[width=\linewidth]{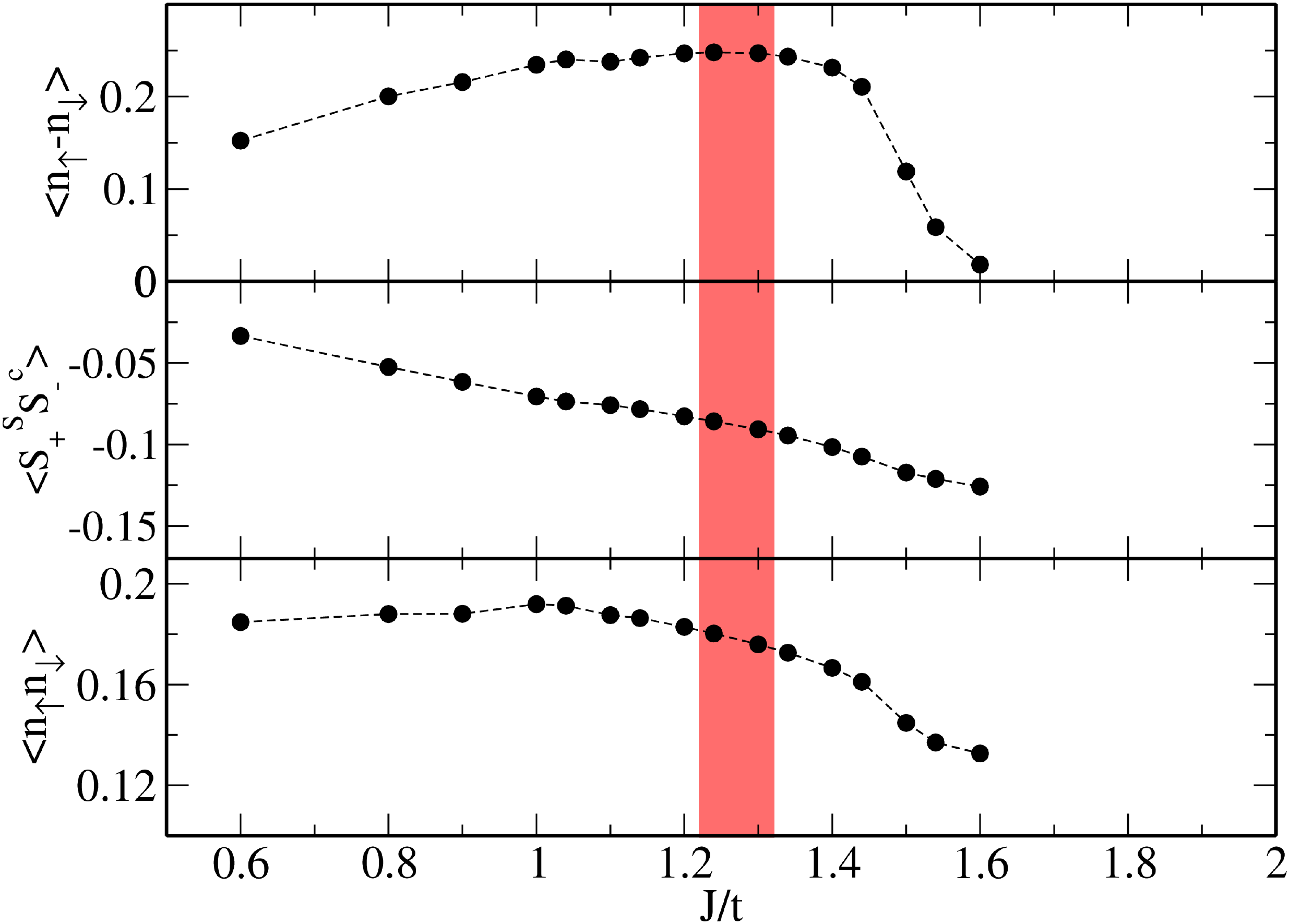}
\end{center}
\caption{ The same quantities as in Fig. \ref{trans_weak} and Fig.
  \ref{trans_strong} across the Lifshitz transition within the SDW phase.
The shaded area corresponds to the interaction strength at
  which the Fermi surface changes from small to large.
\label{trans_J}}
\end{figure}

Finally, let us take a look at expectation values across
the transition of the Fermi surface inside the SDW phase.
Figure \ref{trans_J} shows the same quantities as shown in Fig. \ref{trans_weak}
and Fig. \ref{trans_strong}, but for $\langle n\rangle\approx 0.9$ and
different interaction strengths. We have marked the coupling strengths,
at which the Fermi surface changes from small
to large.
At these interaction strengths, we do not observe any signs of
a transition within the shown expectation values. All quantities
including polarization as well as spin-spin correlations behave smoothly
without any kink. Only for large interaction strength, $J/t\approx
1.6$, we observe the second order phase transition between the SDW
phase with large Fermi surface and the paramagnetic phase. Because
these results are 
for the doped system, this transition occurs already for $J/t\approx
1.6$ compared to $J/t\approx 2.2$ at half filling.

Summarizing our results for the static expectation values, we find a
first order phase transition between the SDW phase with small Fermi
surface at weak coupling and the paramagnetic phase, and a second order phase
transition between the SDW phase with large Fermi surface at strong coupling
and the paramagnetic phase. The change between these two transitions
seems to coincide with the change in the Fermi-surface topology
occurring inside the SDW phase. However, we do not
observe any sign in the static expectation values, although the Fermi surface changes inside the SDW phase.

\section{Conclusions}
We have analyzed the phase diagram of the Kondo lattice model close to
half filling by RDMFT. We have demonstrated the
existence of incommensurate SDW states which coexist with CDWs in
the Kondo lattice model for a wide range of parameters. Although the existence
of SDW states can be expected away from half filling, 
most of the previous
calculations have neglected this possibility.

Remarkably, we have found that within this metallic SDW phase the Fermi surface
changes from small at weak coupling to large at strong coupling. This
Lifshitz transition, which does not seem to affect static quantities
such as polarization and spin-spin correlations, is accompanied
by a change in the order of the phase transition between the SDW phase and
the heavy fermion paramagnetic phase. While we have found a first order
transition between the weak-coupling small Fermi-surface SDW state and
the large Fermi-surface paramagnetic state, we have found a continuous
transition between the large Fermi-surface SDW at strong coupling
and the large-Fermi-surface paramagnetic state.

We want to note that a similar transition between a large Fermi-surface
and a small Fermi-surface antiferromagnetic state has been found in
variational Monte Carlo (VMC)
calculations,\cite{Watanabe2007,Watanabe2008,Lanata2008,Asadzadeh2013} although 
in these calculations incommensurate SDW states have not been
analyzed. These VMC calculations have also shown that the order of the
phase transition between the paramagnetic state and the antiferromagnetic state
changes from first order at weak coupling to second order at strong
coupling. Furthermore, these VMC calculations have confirmed the Lifshitz
transition inside the antiferromagnetic phase. Thus, our RDMFT
calculations, which take incommensurate SDW states into account, agree
with these previous VMC calculations.
Antiferromagnetic states with small and large
  Fermi surface have also been found in a different
  DMFT study\cite{Hoshino2013} and dynamical cluster approximation
  (DCA) studies\cite{Martin2008,Martin2010} using the N\'eel state away from
  half filling instead of SDWs.
Finally, we want to note that a similar Lifshitz transition has been
observed in the ferromagnetic phase of the Kondo lattice
model.\cite{Golez2013}
Thus, a change in the Fermi-surface topology when increasing the
interaction strength seems to be a general
property of the Kondo lattice model and not specific to the SDW phase
close to half filling.

Comparing our results to the global phase diagram for heavy fermions, we
observe two important differences in our results. Although we find 
an SDW phase with small Fermi surface and the other with large Fermi surface,
the phase transition
between the SDW phase with small Fermi surface and the paramagnetic state
is of first order, 
while it is of second order in the global phase
diagram. Second, the spin-spin correlations do not vanish in the SDW
phase with small Fermi surface.
Thus, we do not observe a local quantum critical point in the Kondo
lattice model within the DMFT approximation, which is supposed to be between
the small Fermi-surface SDW phase and the paramagnetic phase. 
This
phase transition turns out to be first order in our calculations.
We only find the previously known continuous quantum phase transition at strong
coupling between the large Fermi-surface SDW and the paramagnetic phase.
 However, we cannot rule out that a local quantum critical
  point might occur in the frustrated KLM, or including long
  range-interactions, or by taking into account spatial fluctuations,
  which are not included in this DMFT study.
By including for example an inter-site exchange, the strength of the
RKKY interaction will be increased, while the Kondo effect is not
strongly influenced. One can thus expect that the antiferromagnetic
phase, and especially the small Fermi-surface antiferromagnetic phase,
extends to larger interactions strengths. Thus, it might be possible
in that case that the Lifshitz transition merges with the continuous
quantum phase transition.

\paragraph*{Acknowledgments ~}
RP thanks for the support through the FPR program of
RIKEN. NK is supported through KAKENHI Grant
No. 25400366. Computer calculations have been done at the
RICC supercomputer at RIKEN and the supercomputer of the Institute of Solid State Physics in Japan.

\bibliography{paper}
\end{document}